\pdfoutput=1
\documentclass[preprint,aps,showpacs,showkeys]{revtex4}

\usepackage{supertabular}
\usepackage{amsfonts}
\usepackage[pdftex]{graphicx}
\usepackage{amsmath}
\usepackage{amsfonts}
\usepackage{amssymb} 
\usepackage{subfig}
\usepackage{amsxtra} 
\newcommand{\be}{\begin{equation}}
\newcommand{\ee}{\end{equation}}
\newcommand{\bn}{\begin{eqnarray}}
\newcommand{\en}{\end{eqnarray}}

\usepackage{amsmath}
\usepackage{amsfonts}
\usepackage{amssymb}
\begin{document}
\title{Properties of fermions with integer spin 
described in Grassmann space~%
\footnote{This contribution developed during
the discussions at the   $20^{\rm{th}}$ --- Bled, 09-17 of July, 2017 ---
 and $21^{\rm{st}} $ --- Bled, 23 of June to 1 of July, 2018 --- Workshops 
"What Comes Beyond the Standard Models", Bled.}}

\author{D. Lukman, N.S. Manko\v c Bor\v stnik${}^1$\\ 
${}^1$University of Ljubljana,\\
Slovenia
}
\begin{abstract}
In Ref.~\cite{nh2018} one of the authors (N.S.M.B.) studies the second quantization of fermions 
with integer spin while describing the internal degrees of freedom of fermions in Grassmann space. 
In this contribution we study the representations in Grassmann space of the groups $SO(5,1)$, 
$SO(3,1)$, $SU(3) \times U(1)$, and $SO(4)$, which are of particular interest as 
the subgroups of the group $SO(13,1)$.
The second quantized integer spin fermions, appearing in Grassmann space, not observed so far,
could be an alternative choice to the half integer spin fermions, appearing in Clifford space. 
The {\it spin-charge-family} theory, using two kinds of Clifford operators --- $\gamma^a$ and 
$\tilde{\gamma}^a$ --- for the description of spins and charges (first) and family quantum numbers 
(second), offers the explanation for not only the appearance of families but also for all the properties of 
quarks and leptons, the gauge fields, scalar fields and others~\cite{norma93,IARD2016,%
n2014matterantimatter,normaJMP2015}. In both cases the gauge fields in $d \ge(13+1)$ --- 
the spin connections $\omega_{ab \alpha}$ (of the two kinds in Clifford case and of 
one kind in Grassmann case) and the vielbeins $f^{\alpha}{}_{\alpha}$ --- determine in $d=(3+1)$ 
scalars, those with
the space index $\alpha=(5,6,\cdots,d)$, and gauge fields, those with the space index $\alpha=(0,1,2,3)$.
While states of the Lorentz group and all its subgroups (in any dimension) are in Clifford space in the 
fundamental representations of the groups, with the family degrees of freedom included~%
\cite{norma93,IARD2016,nh2018}, states in Grassmann space manifest with respect to the
Lorentz group adjoint representations, allowing no families.
\end{abstract}



\maketitle

\section{Introduction}
\label{introduction}

In Ref.~\cite{norma93} the representations in Grassmann  and in Clifford space were discussed.
In Ref.~(\cite{nh2018} and the references therein) the second quantization procedure in both
spaces --- in Clifford space and in Grassmann space --- were discussed in order to try to understand  
"why nature made a choice of Clifford rather than Grassmann space" during the expansion 
of our universe, although in both spaces the creation operators $ \hat{b}^{\dagger}_{j} $  
and the annihilation operators $ \hat{b}_{j} $ exist fulfilling the anticommutation relations 
required for fermions~\cite{nh2018}
\begin{eqnarray}
\{ \hat{b}_i, \hat{b}^{\dagger}_{j} \}_{+}  |\phi_{o}> &=&\delta_{i j}\; |\psi_{o}>\,,
\nonumber\\
\{\hat{b}_i, \hat{b}_{j} \}_{+}  |\psi_{o}>&=& 0\; |\psi_{o}> \,,\nonumber\\
\{\hat{b}^{\dagger}_i,\hat{b}^{\dagger}_{j}\}_{+} |\psi_{o}>
&=&0\; |\psi_{o}> \,,\nonumber\\
\hat{b}^{\dagger}_{j} |\psi_{o}>& =& \, |\psi_{j}>\, \nonumber\\
\hat{b}_{j} |\psi_{o}>& =&0\,  |\psi_{o}>\,.
\label{ijthetaprod} 
\end{eqnarray} 
$|\psi_{o}>$ is  the vacuum state. We use $|\psi_o>= |1>$.
%

 Some observations to be included into introduction:
 However, even in the Grassmann case and with gravity only the scalar fields would appear in  
$d=(3+1)$.

The creation operators can be expressed in both spaces as products of eigenstates 
of the Cartan subalgebra, Eq.~(\ref{choicecartan}), of the Lorentz algebra, 
Eqs.~(\ref{cartaneigengrass}, \ref{signature}). 
Starting with one state (Ref.~\cite{nh2018})
all the other states
of the same representation are reachable  by the  generators of the Lorentz transformations
(which do not belong to the Cartan subalgebra), with ${\cal {\bf S^{ab}}}$ presented in 
Eq.~(\ref{Lorentztheta}) in  Grassmann space and with either $S^{ab}$ or $\tilde{S}^{ab}$, 
Eq.~(\ref{Lorentzgammatilde}), in Clifford space.

 But while there are in Clifford case two kinds of the generators of the Lorentz 
transformations --- $S^{ab}$ and $\tilde{S}^{ab}$, the first transforming members of one 
family among themselves, and the second transforming one member of a particular family into
the same member of other families --- there is in Grassmann space only one kind 
of the Lorentz generators --- ${\cal {\bf S^{ab}}}$. Correspondingly are all the states in 
Clifford space, which can be second quantized as products of nilpotents and projectors~%
\cite{nh02,nh03,nh2018}, 
reachable with one of the two kinds of the operators $S^{ab}$ and $\tilde{S}^{ab}$, 
while different representations are in Grassmann space disconnected. 

On the other hand the vacuum state is in Grassmann case simple --- $|\psi_o>= |1>$ ---
while in Clifford case is the sum of products of projectors, Eq.~(\ref{vac1}). 

In Grassmann space states are in the adjoint representations with respect to the Lorentz 
group, while states in Clifford space belong to the fundamental representations with respect 
to both generators,  $S^{ab}$ and $\tilde{S}^{ab}$, or they are singlets. Correspondingly 
are properties  of fermions, described  with the {\it spin-charge-family} theory~%
\cite{IARD2016,n2014matterantimatter,JMP2013,normaJMP2015,nh2017,nd2017}, which 
uses the Clifford space to describe fermion degrees of freedom, in agreement with the 
observations, offering explanation for all the assumptions of the {\it standard model} 
(with families included) and also other observed phenomena. 

In Grassmann case the spins 
manifest, for example, in the case of $SO(6)$ or $SO(5,1)$ decuplets or singlets --- triplets and 
singlets in Clifford case, 
Table~\ref{Table grassdecupletso51.} --- while with respect to the subgroups $SU(3)$ and 
$U(1)$ of $SO(6)$ the states belong to either singlets, or triplets or sextets, 
Tables~\ref{Table grassdecuplet.},~\ref{Table grasssextet.} 
 --- triplets and singlets in the Clifford case.

In what follows we discuss representations, manifesting as charges and spins of fermions, of 
subgroups of $SO(13,1)$, when internal degrees of freedom of fermions are described in 
Grassmann  space and compare properties of these representations with the properties of 
the corresponding representations appearing in Clifford space. We assume, as in the
{\it spin-charge-family} theory, that both spaces, the internal and the ordinary space, 
have $d=2(2n+1)$-dimensions, $n$ is positive integer, $d \ge 14$ and that all the degrees 
of freedom of fermions and bosons originate in $d=2(2n+1)$, in which fermions interact with 
gravity only. 

After the break of the starting symmetry $SO(13,1)$ into $SO(7,1) \times SU(3) \times U(1)$, and
further to $SO(3,1) \times SU(2) \times SU(2) \times SU(3) \times U(1)$, fermions manifest 
in $d=(3+1)$ the spin and the corresponding charges and interact with the gauge fields, which 
are indeed the spin connections with the space index $m=(0,1,2,3)$, originating in  
$d=(13,1)$~\cite{nd2017}. Also scalar fields originate in gravity: Those spin connections with 
the space index $a =(5,6,7,8)$ determine masses of fermions, those with the space index 
$a=(9,10,\dots,14)$ contribute to particle/antiparticle asymmetry in our 
universe~\cite{n2014matterantimatter}. 

We pay attention in this paper mainly on fermion fields with spin $1$, the creation and 
annihilation operators of which fulfill the anticommutation relations of Eq.~(\ref{ijthetaprod})
in Grassmann space.

\subsection{Creation and annihilation operators in Grassmann space}
\label{grassmann}

In Grassmann $d=2(2n+1)$-dimensional space the creation and annihilation operators follow from
the starting two creation and annihilation operators, both with an odd Grassmann character,  
since those with an even Grassmann character do not obey the anticommutation relations of 
Eq.~(\ref{ijthetaprod})~\cite{nh2018} 
\begin{eqnarray}
\hat{b}^{\theta 1 \dagger}_{1} &=& (\frac{1}{\sqrt{2}})^{\frac{d}{2}} \,
(\theta^0 - \theta^3) (\theta^1 + i \theta^2) (\theta^5 + i \theta^6) \cdots (\theta^{d-1} +
 i \theta^{d}) \,,\nonumber\\
\hat{b}^{\theta 1}_{1} &=& (\frac{1}{\sqrt{2}})^{\frac{d}{2}}\,
 (\frac{\partial}{\;\partial \theta^{d-1}} - i \frac{\partial}{\;\partial \theta^{d}})
\cdots (\frac{\partial}{\;\partial \theta^{0}}
-\frac{\partial}{\;\partial \theta^3})\,,\nonumber\\
\hat{b}^{\theta 2 \dagger }_{1} &=& (\frac{1}{\sqrt{2}})^{\frac{d}{2}} \,
(\theta^0 + \theta^3) (\theta^1 + i \theta^2) (\theta^5 + i \theta^6) \cdots (\theta^{d-1} +
 i \theta^{d}) \,,\nonumber\\
\hat{b}^{\theta 2}_{1} &=& (\frac{1}{\sqrt{2}})^{\frac{d}{2}}\,
 (\frac{\partial}{\;\partial \theta^{d-1}} - i \frac{\partial}{\;\partial \theta^{d}})
\cdots (\frac{\partial}{\;\partial \theta^{0}}
+\frac{\partial}{\;\partial \theta^3})\,.
\label{start(2n+1)2theta}
\end{eqnarray}
All the creation operators are products of the eigenstates of the Cartan subalgebra operators,%
Eq.~(\ref{choicecartan})
\begin{eqnarray}
\label{cartaneigengrass}
{\cal {\bf S}}^{ab} (\theta^a \pm \epsilon \theta^b) &=& \mp i  
\frac{\eta^{aa}}{\epsilon} (\theta^a \pm \epsilon \theta^b)\,, \nonumber\\
\epsilon = 1\,, \;\, {\rm for}\;\, \eta^{aa}=1\,,&&
\epsilon = i \,,\;\, {\rm for}\;\, \eta^{aa}= -1\,, \nonumber\\
{\cal {\bf S}}^{ab}\, (\theta^a \theta^b \pm \epsilon \theta^c \theta^d)=0\,,&& 
 {\cal {\bf S}}^{cd}\, (\theta^a \theta^b \pm \epsilon \theta^c \theta^d)=0\,.
\end{eqnarray}

The two creation operators, $\hat{b}^{\theta 1 \dagger}_{1}$ and 
$\hat{b}^{\theta 2 \dagger}_{1}$, if applied on the vacuum state, form the starting two states 
$\phi^{1}_{1}$ and $\phi^{2}_{1}$ of the two representations, respectively. The vacuum state 
is chosen to be the simplest one~\cite{nh2018} --- $|\phi_{0}> = |1>$. The rest of creation operators
of each of the two groups, $\hat{b}^{\theta 1 \dagger}_{i}$ and $\hat{b}^{\theta 2 \dagger}_{i}$, 
follow from the starting one by the application of the generators of the Lorentz transformations in 
Grassmann space ${\cal {\bf S}}^{ab}$, Eq.~(\ref{Lorentztheta}), which do not belong to the 
Cartan subalgebra, Eq.~(\ref{choicecartan}), of the Lorentz algebra. They generate either 
$|\phi^{1}_{j}>$ of the first group or $|\phi^{2}_{j}>$ of the second group.

Annihilation operators $\hat{b}^{\theta 1}_{i}$ and $\hat{b}^{\theta 2}_{i}$ follow from the 
creation ones by the Hermitian conjugation~\cite{nh2018}, when taking into account the 
assumption
\begin{eqnarray}
\label{grassher}
(\theta^a)^{\dagger} &=&  \frac{\partial}{\partial \theta_{a}} \eta^{aa}=
-i \,p^{\theta a} \eta^{aa}\,, 
\end{eqnarray}
from where it follows
\begin{eqnarray}
\label{grassp}
(\frac{\partial}{\partial \theta_{a}})^{\dagger} &=& \eta^{aa}\, \theta^a\,,\quad
(p^{\theta a})^{\dagger} = -i \eta^{aa} \theta^a\,.
\end{eqnarray}

The annihilation operators $\hat{b}^{\theta 1}_{i}$ and $\hat{b}^{\theta 2}_{i}$ annihilate 
states $|\phi^{1}_{i}>$ and $|\phi^{2}_{i}>$, respectively.  

The application of ${\cal {\bf S}}^{01}$ on $\hat{b}^{\theta 1 \dagger}_{1}$, for example,
 transforms this creation operator into $\hat{b}^{\theta 1 \dagger}_{2} = $ 
$(\frac{1}{\sqrt{2}})^{\frac{d}{2} -1} \,(\theta^0  \theta^3 +i \theta^1 \theta^2)$
$ (\theta^5 + i \theta^6) \cdots (\theta^{d-1} - i \theta^{d})$. Correspondingly its Hermitian 
conjugate annihilation operator is equal to
$\hat{b}^{\theta 1}_{2} = (\frac{1}{\sqrt{2}})^{\frac{d}{2}-1}\,
 (\frac{\partial}{\;\partial \theta^{d-1}} - i \frac{\partial}{\;\partial \theta^{d}})
\cdots (\frac{\partial}{\;\partial \theta^{3}} \,\frac{\partial}{\;\partial \theta^0} - i 
\frac{\partial}{\;\partial \theta^{2}} \,\frac{\partial}{\;\partial \theta^1})$.

All the states are normalized with respect to the integral over the Grassmann coordinate  
space~\cite{norma93}

\begin{eqnarray}
\label{grassnorm}
<\phi^{a}_{i}|\phi^{b}_{j} > &=&  \int d^{d-1} x  d^d \theta^a\, \,\omega 
 <\phi^{a}_{i}|\theta> <\theta|\phi^{b}_{j} > = \delta^{ab}\,\delta_{ij} \,, \nonumber\\
\omega&=& \Pi^{d}_{k=0}(\frac{\partial}{\;\,\partial \theta_k} + \theta^{k})\,,
\end{eqnarray}
where $\omega$ is a weight function, defining the scalar product $<\phi^a_{i}|\phi^b_{j} >$,
 and we require that~\cite{norma93} 
\begin{eqnarray}
\label{grassintegral}
\{ d\theta^a, \theta^b \}_{+} &=&0, \,\;\;  \int d\theta^a  =0\,,\,\;\; 
\int d\theta^a \theta^a =1\,,\nonumber\\
\int d^d \theta \,\,\theta^0 \theta^1 \cdots \theta^d &=&1\,,
\nonumber\\
d^d \theta &=&d \theta^d \dots d\theta^0\,,
\end{eqnarray}
with $ \frac{\partial}{\;\,\partial \theta_a} \theta^c = \eta^{ac}$. 

There are $\frac{1}{2}\, \frac{d!}{\frac{d!}{2} \frac{d!}{2}}$ in each of these two groups 
of creation operators of an odd Grassmann character in $d=2(2n+1)$-dimensional space.

The rest of creation operators (and the corresponding annihilation operators) would have  rather 
opposite  Grassmann character than the ones studied so far: like {\bf a.} $\theta^0 \theta^1$ 
for the creation operator and 
[$\frac{\partial}{\partial \theta^1}\frac{\partial}{\partial \theta^0}$]  for the 
corresponding annihilation operator in $d=(1+1)$
 (since $\{\theta^0 \theta^1, \frac{\partial}{\partial \theta^{1}}$ 
 $\frac{\partial}{\partial \theta^{0}} \}_{+}$ gives $(1+ (1+1) \theta^0 \theta^1 $
 $\frac{\partial}{\partial \theta^{1}} \frac{\partial}{\partial \theta^{0}}) $), and like {\bf b.}
 $(\theta^0 \mp \theta^3) (\theta^1\pm i \theta^2)$ for creation operator and 
[$ (\frac{\partial}{\partial \theta^1}\mp i \frac{\partial}{\partial \theta^2}) 
(\frac{\partial}{\partial \theta^0} \mp \frac{\partial}{\partial \theta^3})$] for the
annihilation operator,   or $\theta^0 \theta^3
 \theta^1  \theta^2$ for the creation operator and [$\frac{\partial}{\partial \theta^2} 
\,\frac{\partial}{\partial \theta^1}\,\frac{\partial}{\partial \theta^3}\,
\frac{\partial}{\partial \theta^0}$] for the annihilation operator in $d=(3+1)$ 
 (since, let say, $\{\frac{1}{2} (\theta^0 - \theta^3) (\theta^1 + i \theta^2),$ 
 $\frac{1}{2} (\frac{\partial}{\partial \theta^1} - i \frac{\partial}{\partial \theta^2}) 
 (\frac{\partial}{\partial \theta^0} - \frac{\partial}{\partial \theta^3})\}_{+}$ gives
 $(1 + \frac{1}{4} (1+1)  (\theta^0 - \theta^3) (\theta^1 + i \theta^2)  
 (\frac{\partial}{\partial \theta^1} - i \frac{\partial}{\partial \theta^2}) 
 (\frac{\partial}{\partial \theta^0} - \frac{\partial}{\partial \theta^3})$ and equivalently for 
other  cases), but applied on a vacuum states some of them still fulfill some of the relations
 of Eq.~(\ref{ijthetaprod}), but not all (like $\{\frac{1}{2}(\theta^0 - \theta^3) 
(\theta^1 + i \theta^2),$  $\frac{1}{2} (\theta^0 + \theta^3) (\theta^1 - i \theta^2)\}_{+}
=$  $i \theta^0 \theta^1\theta^2 \theta^3 $, while it should be zero). 

Let us add that, like in Clifford case, one can simplify the scalar product in Grassmann case 
by recognizing that the scalar product  is equal to $\delta^{ab}\,\delta_{ij}$
\begin{eqnarray}
\label{grassscalar}
<\phi^{a}_{i}|\theta> <\theta|\phi^{b}_{j} > &=& \delta^{ab}\,\delta_{ij}\,,
\end{eqnarray}
without  integration over the Grassmann coordinates. Let us manifest this in the case
of $d=(1+1)$:$<1|\frac{1}{\sqrt{2}}(\frac{\partial}{\partial \theta^0} - 
\frac{\partial}{\partial \theta^1})
\frac{1}{\sqrt{2}}(\theta^0 - \theta^1)|1>=1 $, $|1>$ is the normalized vacuum state,
$<1|1>=1$. It is true 
in all dimensions, what can easily be understood for all the states, which are defined by 
the creation operators $\hat{b}_{i}^{\dagger}$ on the vacuum state $|1>$, 
$|\phi^{b}_{i} >= \hat{b}_{i}^{\dagger}|1>$, fulfilling the anticommutation relations
of Eq.~(\ref{ijthetaprod}).

 
%
\subsection{Creation and annihilation operators in Clifford space}
\label{clifford}

There are two kinds of Clifford objects~\cite{norma93}, (\cite{IARD2016} and Refs. therein),
$\gamma^a$ and $\tilde{\gamma}^a$, both fulfilling the anticommutation relations
\begin{eqnarray}
\label{tildecliffcomrel}
 \{\gamma^a, \gamma^b \}_{+} &=& 2 \eta^{a b} = 
\{\tilde{\gamma}^a, \tilde{\gamma}^b \}_{+}\,, \nonumber\\ 
 \{\gamma^a, \tilde{\gamma}^b \}_{+}&=&0\,. 
\end{eqnarray}
Both Clifford algebra objects are expressible with $\theta^a $ and 
$\frac{\partial}{\,\;\partial \theta^a}$~\cite{norma93,nh2018}, 
 (\cite{IARD2016} and Refs. therein)
\begin{eqnarray}
\label{cliffthetarel}
\gamma^a &=& (\theta^a + \frac{\partial}{\;\partial \theta_a})\,,\nonumber\\
\tilde{\gamma}^a &=& i \, (\theta^a - \frac{\partial}{\;\partial \theta a})\,, \nonumber\\
\theta^a &=&\frac{1}{2} (\gamma^a - i \tilde{\gamma}^a)\,,\nonumber\\
\frac{\,\partial}{\partial \theta_a} &=&\frac{1}{2} \, (\gamma^a + i \tilde{\gamma}^a)\,,
\end{eqnarray}
from where it follows: $(\gamma^a)^{\dagger} = \gamma^a \eta^{aa}$, 
$(\tilde{\gamma}^a)^{\dagger} = \tilde{\gamma}^a \eta^{aa}$,
$\gamma^a \gamma^a = \eta^{aa}$, $\gamma^a (\gamma^a)^{\dagger} =1$,
$ \tilde{\gamma}^a  \tilde{\gamma}^a = \eta^{aa}$, 
$ \tilde{\gamma}^a  (\tilde{\gamma}^a)^{\dagger} =1$.

Correspondingly we can use either $\gamma^a$  or  $\tilde{\gamma}^a$ instead of 
$\theta^a$ to span the internal space of fermions. Since both, $\gamma^a$ and 
$\tilde{\gamma}^a$, are expressible with $\theta^a$ and the derivatives with respect to 
$\theta^a$, the norm of vectors in Clifford space can be defined by the same integral as in 
Grassmann space, Eq.(\ref{grassnorm}), or we can simplify the scalar product (as in the 
Grassmann case, Eq.~(\ref{grassscalar}) by introducing
the Clifford vacuum state $|\psi_{oc}>$, Eq.~(\ref{vac1}), instead of $|1>$ in Grassmann
case. 

We make use of $\gamma^a$ to span the vector space. As in the case of Grassmann
space we require that the basic states are eigenstates of the Cartan subalgebra operators of 
$S^{ab}$ and $\tilde{S}^{ab}$, Eq.~(\ref{choicecartan}). 
\begin{eqnarray}
\stackrel{ab}{(k)}:&=& 
\frac{1}{2}(\gamma^a + \frac{\eta^{aa}}{ik} \gamma^b)\,,\quad 
\stackrel{ab}{(k)}^{\dagger} = \eta^{aa}\stackrel{ab}{(-k)}\,,\nonumber\\
\stackrel{ab}{[k]}:&=&
\frac{1}{2}(1+ \frac{i}{k} \gamma^a \gamma^b)\,,\quad \;\,
\stackrel{ab}{[k]}^{\dagger} = \,\stackrel{ab}{[k]}\,,\nonumber\\
S^{ab}\,\stackrel{ab}{(k)} &=& \frac{1}{2} k\, \stackrel{ab}{(k)}\,,\quad \quad \quad
S^{ab}\,\stackrel{ab}{[k]}  = \frac{1}{2} k\, \stackrel{ab}{[k]}\,, \nonumber\\
\tilde{S}^{ab}\,\stackrel{ab}{(k)} &=& \frac{1}{2} k\, \stackrel{ab}{(k)}\,,\quad \quad \quad
\tilde{S}^{ab}\,\stackrel{ab}{[k]}  = -\frac{1}{2} k\, \stackrel{ab}{[k]}\,,
\label{signature}
\end{eqnarray}
with $k^2 = \eta^{aa} \eta^{bb}$. To calculate $\tilde{S}^{ab}\,\stackrel{ab}{(k)}$ and 
$\tilde{S}^{ab}\,\stackrel{ab}{[k]}$ we define~\cite{nh03,nh02} the application of 
$\tilde{\gamma}^a$  on any Clifford algebra object A  as follows
\begin{eqnarray}
(\tilde{\gamma^a} A = i (-)^{(A)} A \gamma^a)|\psi_{oc}>\,,
\label{gammatildeA}
\end{eqnarray}
where $A$ is any Clifford algebra object and  $(-)^{(A)} = -1$,  if $A$ is an odd Clifford algebra
 object and $(-)^{(A)} = 1$,  if $A$  is an even Clifford algebra
 object, $|\psi_{oc}>$ is the vacuum state, replacing  the vacuum state $|\psi_o>= |1>$,
 used in Grassmann case, with the 
one  of Eq.~(\ref{vac1}), in accordance with the relation of Eqs.~({\ref{cliffthetarel},
\ref{grassnorm}, \ref{grassintegral}}), Ref.~\cite{nh2018}.

We can define now the creation and annihilation operators in Clifford space so that they fulfill the
requirements of Eq.~(\ref{ijthetaprod}).
We write the starting creation operator and its Hermitian conjugate one (in 
accordance with Eq.~(\ref{signature}) and Eq.(\ref{choicecartan})) in $2(2n+1)$-dimensional 
space as follows~\cite{nh2018}
\begin{eqnarray}
\hat{b}^{1 \dagger}_1&=& \stackrel{03}{(+i)} \stackrel{12}{(+)} \stackrel{56}{(+)}\cdots
\stackrel{d-1\;d}{(+)}\,,\nonumber\\
\hat{b}^{1}_1&=& \stackrel{d-1\;d}{(-)} \cdots \stackrel{56}{(-)} \stackrel{12}{(-)}
\stackrel{03}{(-i)}\,.
\label{bstart}
\end{eqnarray}
The starting creation operator $\hat{b}^{1  \dagger}_1$, when applied on the vacuum state
$|\psi_{oc}>$, defines the starting family member of the starting ''family". The corresponding
starting annihilation operator is its Hermitian conjugated one, Eq.~(\ref{signature}).

All the other creation operators of the same family can be obtained by the application of the 
generators of the Lorentz transformations $S^{ab}$, Eq.~(\ref{Lorentzgammatilde}), which do 
not belong to the Cartan subalgebra of $SO(2(2n+1) -1,1)$, Eq.~(\ref{choicecartan}). 
\begin{eqnarray}
\hat{b}^{1\dagger}_i &\propto & S^{ab} ..S^{ef} \hat{b}^{1\dagger}_1\,,\nonumber\\
\hat{b}^{1}_i&\propto & \hat{b}^{1}_1 S^{ef}..S^{ab}\,,
\label{b1i}
\end{eqnarray}
with $S^{ab\dagger} = \eta^{aa} \eta^{bb} S^{ab}$. The proportionality factors are chosen 
so, that the corresponding states $|\psi^{1}_{1}>= \hat{b}^{1\dagger}_i |\psi_{oc}>$ are 
normalized, where $|\psi_{oc}>$ is the normalized vacuum state, $<\psi_{oc}|\psi_{oc}> =1$. 

The creation operators creating different "families" with respect to the starting "family",
 Eq.~(\ref{bstart}), 
can be obtained from the starting one by the application of $\tilde{S}^{ab}$, 
Eq.~(\ref{Lorentzgammatilde}), which do not belong to the Cartan subalgebra of 
$\widetilde{SO}(2(2n+1) -1,1)$, Eq.~(\ref{choicecartan}). They all keep the "family member" 
quantum number unchanged.
\begin{eqnarray}
\hat{b}^{\alpha \dagger}_i &\propto & \tilde{S}^{ab} \cdots \tilde{S}^{ef}\,
\hat{b}^{1\dagger}_{i}\,.
\label{balpha1}
\end{eqnarray}
Correspondingly we can define (up to the proportionality factor) any creation operator for any
"family" and any "family member" with the application of $S^{ab}$ and $\tilde{S}^{ab}$%
~\cite{nh2018}
\begin{eqnarray}
\hat{b}^{\alpha \dagger}_i&\propto & \tilde{S}^{ab} \cdots \tilde{S}^{ef} 
{S}^{mn}\cdots {S}^{pr}
\hat{b}^{1\dagger}_{1}\nonumber\\
&\propto & {S}^{mn}\cdots {S}^{pr} \hat{b}^{1\dagger}_{1} {S}^{ab} \cdots {S}^{ef}\,.
\label{anycreation}
\end{eqnarray}
All the corresponding annihilation operators follow from the creation ones by the Hermitian 
conjugation.

There are $2^{\frac{d}{2}-1}$ $\times \;\, 2^{\frac{d}{2}-1}$ creation operators of an odd 
Clifford character and the same number of annihilation operators, which fulfill the anticommutation
relations of Eq.~(\ref{ijthetaprod}) on the vacuum state $|\psi_{oc}>$ with
$2^{\frac{d}{2}-1}$ summands
\begin{eqnarray}
|\psi_{oc}>&=& \alpha\,( \stackrel{03}{[-i]} \stackrel{12}{[-]} 
\stackrel{56}{[-]}\cdots
\stackrel{d-1\;d}{[-]} + \stackrel{03}{[+i]} \stackrel{12}{[+]} \stackrel{56}{[-]} \cdots
\stackrel{d-1\;d}{[-]} + \stackrel{03}{[+i]} \stackrel{12}{[-]} \stackrel{56}{[+]}\cdots
\stackrel{d-1\;d}{[-]} + \cdots ) |0>\,, \quad \nonumber\\
&&\alpha =\frac{1}{\sqrt{2^{\frac{d}{2}-1}}}\,, \nonumber\\
&&{\rm for}\; d=2(2n+1)\,,
\label{vac1}
\end{eqnarray}
$n$ is a positive integer. For a chosen $\alpha =\frac{1}{\sqrt{2^{\frac{d}{2}-1}}}$
the vacuum is normalized: $<\psi_{oc}|\psi_{oc}>=1$.

It is proven in Ref.~\cite{nh2018} that the creation and annihilation operators fulfill the
anticommutation relations required for fermions, Eq.~(\ref{ijthetaprod}).

\section{Properties of representations of the Lorentz group $SO(2(2n+1))$ and of subgroups
in Grassmann and in Clifford space}
\label{grassmanncliffordrepresentations}

The purpose of this contribution is to compare properties of the representations of the Lorentz
group $SO(2(2n+1))$,  $n \ge 3$, when for the description of the internal degrees of freedom 
of fermions either {\bf i.}  Grassmann space or {\bf ii.} Clifford space is used. The 
{\it spin-charge-family} theory~(\cite{JMP2013,normaJMP2015,IARD2016,%
n2014matterantimatter,nh2017,nd2017,n2012scalars} and the references therein) namely
predicts that all the properties of the observed either quarks and leptons or vector gauge fields 
or scalar gauge fields originate in $d \ge (13+1)$, in which massless fermions interact with the 
gravitational field only --- with its spin connections and 
vielbeins. 

However, both --- Clifford space and Grassmann space --- allow second quantized states, the 
creation and annihilation operators of which fulfill the anticommutation relations for fermions of 
Eq.~(\ref{ijthetaprod}). 

But while Clifford space offers the description of spins, charges and families of fermions in 
$d=(3+1)$, all in the fundamental representations of the Lorentz group $SO(13,1)$  and the 
subgroups of the Lorentz group, in agreement with the observations, the representations of the 
Lorentz group are in Grassmann space the adjoint ones, in disagreement with what we  
observe.

We compare properties of the representations in Grassmann case with those in Clifford case 
to be able  to better understand "the choice of nature in the expanding universe, making use of the 
Clifford degrees of freedom", rather than Grassmann degrees of freedom.

In introduction we briefly reviewed properties of creation and annihilation operators in both spaces,
presented in Ref.~\cite{nh2018} (and the references therein). 
We pay attention on spaces with $d=2(2n+1)$ of ordinary coordinates and  $d=2(2n+1)$ internal 
coordinates, either of Clifford or of Grassmann character. 

{\bf i.} $\;\;$ In Clifford case there are $2^{\frac{d}{2} - 1}$ creation operators of an odd 
Clifford character, creating "family members" when applied on the vacuum state. We choose 
them to be eigenstates of the Cartan subalgebra operators, Eq.(\ref{choicecartan}), of the Lorentz 
algebra. All the members
can be reached from any of the creation operators  by the application of $S^{ab}$, 
Eq.~(\ref{Lorentzgammatilde}).
Each "family member" appears in $2^{\frac{d}{2} - 1}$ "families", again of an odd Clifford 
character, since the corresponding creation operators are  reachable by $\tilde{S}^{ab}$, 
Eq.~(\ref{Lorentzgammatilde}), which are Clifford even objects. 

There are correspondingly $2^{\frac{d}{2} - 1} \cdot$ $2^{\frac{d}{2} - 1}$ creation and the 
same number ($2^{\frac{d}{2} - 1} \cdot$ $2^{\frac{d}{2} - 1}$) of annihilation operators. Also 
the annihilation operators, annihilating states of $2^{\frac{d}{2} - 1}$ "family members" in
 $2^{\frac{d}{2} - 1}$ "families", have an odd Clifford character, since they are Hermitian conjugate
 to the creation ones. 

The rest of $2 \cdot$ $2^{\frac{d}{2} - 1}\cdot$ $2^{\frac{d}{2} - 1}$ members
of the Lorentz representations  have an even Clifford character, what means that the corresponding 
creation and annihilation operators can not fulfill the anticommutation relations required for 
fermions, Eq.~(\ref{ijthetaprod}). 
Among these $2^{\frac{d}{2} - 1}$ products of projectors determine the vacuum state, 
Eq.~(\ref{vac1}).


{\bf ii.} $\;\;$ In Grassmann case there are $\frac{d!}{\frac{d}{2}!\,\frac{d}{2}! }$ 
operators of an odd
Grassmann character, which form the creation operators, fulfilling with the corresponding 
annihilation operators the requirements of  Eq.~(\ref{ijthetaprod}). All the creation operators 
are chosen to be products of the eigenstates of the Cartan subalgebra ${\cal {\bf S}}^{ab}$, 
Eq.~(\ref{choicecartan}).  The corresponding annihilation operators are the Hermitian conjugated 
values of the creation operators, Eqs.~(\ref{grassher}, \ref{grassp}, \ref{start(2n+1)2theta}). 
The creation operators form, when applied on the simple vacuum state $|\phi_{o}> = |1>$,
two independent groups of states. The members of each of the two groups are 
reachable from any member of a group by the application of ${\cal {\bf S}}^{ab}$, 
Eq.~(\ref{Lorentztheta}). All the states of any of the two decuplets are orthonormalized.

We comment in what follows the representations in $d=(13+1)$ in Clifford and in Grassmann case.
In {\it spin-charge family} theory there are breaks of the starting symmetry from $SO(13,1)$ to
$SO(3,1)\times SU(2) \times SU(3) \times U(1)$ in steps, which lead to the so far observed 
quarks and leptons, gauge and scalar fields and gravity. One 
of the authors (N.S.M.B.), together with H.B. Nielsen, defined the discrete symmetry operators for 
Kaluza-Klein theories for spinors in Clifford space~\cite{nhds}. In Ref.~\cite{nh2018} the same 
authors define the discrete 
symmetry operators in the case that for the description of fermion degrees of 
freedom Grassmann space is used. Here we comment symmetries in both spaces for some of
subgroups of the $SO(13,1)$ group, as well as the appearance of the Dirac sea.

\subsection{Currents in Grassmann space} 
\label{currents}
%

%
\begin{eqnarray}
\{ \theta^a p_a, \frac{\partial }{\partial \theta_b} p_b \}_{+}&=&
 \theta^a p_a  \frac{\partial }{\partial \theta_b} p_b + 
 \frac{\partial }{\partial \theta_b} p_b \theta^a p_a = \eta^{ab}p_a p_b \,, 
\label{KGgrass}
\end{eqnarray}
\begin{eqnarray}
\phi^{\dagger}(\theta^0 + \frac{\partial }{\partial \theta_0})
(\theta^a +  \frac{\partial }{\partial \theta_a}) \phi &=& j^{a} \,, 
\label{currentgrass}
\end{eqnarray}
%

%
\subsection{Equations of motion in Grassmann and Clifford space}
\label{equationinCandG}

We define~\cite{nh2018} the action in Grassmann  space, for which we require --- similarly as in 
 Clifford case --- that the action for a free massless object 
%
\begin{eqnarray}
{\cal A}\,  &=& \frac{1}{2}  \int \; d^dx \;d^d\theta\; \omega \, \{\phi^{\dagger} 
(1-2\theta^0 \frac{\partial}{\partial \theta^0}) \,\frac{1}{2}\,
 (\theta^a p_{a}+ \eta^{aa} \theta^{a \dagger}
 p_{a}) \phi \} \,, 
\label{actionWeylGrass}
\end{eqnarray}
is Lorentz invariant. The corresponding equation of motion is 
\begin{eqnarray}
\label{Weylgrass}
\frac{1}{2}[ (1-2\theta^0 \frac{\partial}{\partial \theta^0}) \,\theta^a + 
((1-2\theta^0 \frac{\partial}{\partial \theta^0}) \,\theta^a)^{\dagger}]\,
\, p_{a} \,|\phi^{\theta}_{i}>\,&= & 0\,,
\end{eqnarray}
$p_{a} = i\, \frac{\partial}{\partial x^a}$, leading to the Klein-Gordon equation
\begin{eqnarray}
\label{LtoKGgrass}
\{(1-2\theta^0 \frac{\partial}{\partial \theta^0}) \theta^a p_{a}\}^{\dagger}\,\theta^b p_{b}
|\phi^{\theta}_{i}>&= &  
p^a p_a |\phi^{\theta}_{i}>=0\,.
\end{eqnarray}
In the Clifford case the action for massless fermions is well known
\begin{eqnarray}
{\cal A}\,  &=& \int \; d^dx \; \frac{1}{2}\, (\psi^{\dagger}\gamma^0 \, \gamma^a p_{a} \psi) +
 h.c.\,, 
\label{actionWeyl}
\end{eqnarray}
 leading to the equations of motion 
\begin{eqnarray}
\label{Weyl}
\gamma^a p_{a}  |\psi^{\alpha}>&= & 0\,, 
\end{eqnarray}
which fulfill also the Klein-Gordon equation
\begin{eqnarray}
\label{LtoKG}
\gamma^a p_{a} \gamma^b p_b |\psi^{\alpha}_{i}>&= &   
p^a p_a |\psi^{\alpha}_{i}>=0\,.
\end{eqnarray}
\subsection{Discrete symmetries in Grassmann and Clifford space}
\label{CPT}

We follow also here Ref.~\cite{nh2018} and the references therein.
We distinguish in $d$-dimensional space two kinds of dicsrete operators ${\cal C}, {\cal P}$ 
and ${\cal T}$
 operators with respect to the internal space which we use.

In the Clifford case~\cite{nhds}, when the whole $d$-space is treated equivalently, we have 
\begin{eqnarray}
\label{calCPTH}
{\cal C}_{{\cal H}}&=& \prod_{\gamma^a \in \Im} \gamma^a \,\, K\,,\quad
{\cal T}_{{\cal H}}= \gamma^0 \prod_{\gamma^a \in \Re} \gamma^a \,\, K\, I_{x^0}\,\,\,,
 \quad {\cal P}^{(d-1)}_{{\cal H}} = \gamma^0\,I_{\vec{x}}\,,\nonumber\\
I_{x} x^a &=&- x^a\,, \quad I_{x^0} x^a = (-x^0,\vec{x})\,, \quad I_{\vec{x}} \vec{x} =
 -\vec{x}\,, \nonumber\\
I_{\vec{x}_{3}} x^a &=& (x^0, -x^1,-x^2,-x^3,x^5, x^6,\dots, x^d)\,.
\end{eqnarray}
The product $\prod \, \gamma^a$ is meant in the ascending order in $\gamma^a$.

In the Grassmann case we correspondingly define
\begin{eqnarray}
\label{calCPTG}
{\cal C}_{G}&=& \prod_{\gamma^a_{ G} \in \Im \gamma^a} \, \gamma^a_{ G}\, K\,,\quad
{\cal T}_{G} = \gamma^0_{G} \prod_{\gamma^a_{G} \in \Re
 \gamma^a} \,\gamma^a_{ G}\, K \, I_{x^0}\,,\quad
{\cal P}^{(d-1)}_{G} = \gamma^0_{G} \,I_{\vec{x}}\,,
\end{eqnarray}
with $\gamma^a_{G}$  defined as 
\begin{eqnarray}
\label{gammaG}
\gamma^{a}_{G} &=& 
(1- 2 \theta^a \eta^{aa} \frac{\partial}{\partial \theta_{a}})\,, 
\end{eqnarray}
while $I_{x}$,
$I_{\vec{x}_{3}}$ 
is defined in Eq.~(\ref{calCPTH}).
Let be noticed, that since $\gamma^a_{G}$ ($= - i \eta^{aa}\, \gamma^a \tilde{\gamma}^a$) is 
always real as there is $ \gamma^a i \tilde{\gamma}^a$, while $ \gamma^a$ is either real or 
imaginary,
we use in Eq.~(\ref{calCPTG}) $\gamma^a$  to make a choice of appropriate $\gamma^a_{G}$. 
In what follows we shall use the notation as in Eq.~(\ref{calCPTG}).

We define, according to Ref.~\cite{nh2018} (and the references therein)  in both cases --- Clifford 
Grassmann case --- the operator
 "emptying"~\cite{JMP2013,normaJMP2015} (arxiv:1312.1541) the Dirac sea, so that operation of 
"emptying$_{N}$" after the charge conjugation ${\cal C }_{{\cal H}}$ in the Clifford case and 
"emptying$_{G}$" after the charge conjugation ${\cal C }_{G}$
in the Grassmann case (both transform the state put on the top of either the Clifford or the Grassmann
Dirac sea into the corresponding negative energy state) creates the anti-particle state to the starting 
particle state, both  put on the top of the Dirac sea and both solving the Weyl equation, either in the
Clifford case, Eq.~(\ref{Weyl}),  or in the Grassmann case, Eq.~(\ref{Weylgrass}), for free massless 
fermions
\begin{eqnarray}
\label{empt}
"{\rm emptying}_{N}"&=& \prod_{\Re \gamma^a}\, \gamma^a \,K\, \quad {\rm in} \, \;
{\rm Clifford}\, {\rm space}\,, 
\nonumber\\
"{\rm emptying}_{G}"&=& \prod_{\Re \gamma^a}\, \gamma^a_{G} \,K\, \quad {\rm in}\;\, 
{\rm Grassmann} \,  {\rm space}\,, 
\end{eqnarray}
although we must keep in mind that indeed the anti-particle state is a hole in the Dirac sea from the 
Fock space point of view. The operator "emptying" is bringing the single particle operator 
${\cal C }_{{\cal H}}$ in the Clifford case and ${\cal C }_{G}$ in the Grassmann case into the operator 
on the Fock space in each of the two cases.
Then the anti-particle  state creation operator --- 
${\underline {\bf {\Huge \Psi}}}^{\dagger}_{a}[\Psi_{p}]$ --- to the corresponding  particle state 
creation operator --- can be obtained also as follows
\begin{eqnarray}
\label{makingantip}
{\underline {\bf {\Huge \Psi}}}^{\dagger}_{a}[\Psi_{p}]\, |vac>  &=& 
{\underline {\bf \mathbb{C}}}_{{{\bf \cal H}}}\, 
{\underline {\bf {\Huge \Psi}}}^{\dagger}_{p}[\Psi_{p}]\, |vac>  =  
\int \, {\mathbf{\Psi}}^{\dagger}_{a}(\vec{x})\, 
({\bf \mathbb{C}}_{\cal H}\,\Psi_{p} (\vec{x})) \,d^{(d-1)} x  \, \,|vac> \,,\nonumber\\
{\bf \mathbb{C}}_{\cal H} &=& "{\rm emptying}_{N}"\,\cdot\, {\cal C}_{{\cal H}}  \,
\end{eqnarray}
in both cases.

The operators ${\bf \mathbb{C}}_{\cal H}$ and ${\bf \mathbb{C}}_{G}$ 
\begin{eqnarray}
\label{emptCHG}
{\bf \mathbb{C}}_{\cal H} &=& "{\rm emptying}_{N}" \,\cdot\, {\cal C}_{{\cal H}} \,,\nonumber\\
{\bf \mathbb{C}}_{G} &=& "{\rm emptying}_{NG}" \,\cdot\, {\cal C}_{G}\,,
\end{eqnarray}
 operating on 
$\Psi_{p} (\vec{x})$ transforms the positive energy spinor state (which solves the corresponding 
Weyl equation for a massless free fermion) put on the top of the Dirac sea into the positive energy 
anti-fermion state, which again solves the corresponding Weyl equation for a massless free 
anti-fermion put on the top of the Dirac sea. Let us point out that either the operator 
$"{\rm emptying}_{N}" $ 
or the operator $"{\rm emptying}_{NG}"$ transforms the single particle operator either
$ {\cal C}_{\cal H}$ or ${\cal C}_{G}$ into the operator operating in the Fock space.

We use the Grassmann even, Hermitian and real operators $\gamma^{a}_{G}$, 
Eq.~(\ref{gammaG}), to define discrete symmetry in Grassmann space, first we did in 
$((d+1)-1)$ space, Eq.~(\ref{calCPTG}), now we do in $(3+1)$  space, Eq.~(\ref{calCPTNG}),
 as it is done 
in~\cite{nhds} in the Clifford case. 
In the Grassmann case we do this in analogy with the operators in the Clifford case~\cite{nhds}
\begin{eqnarray}
\label{calCPTNG}
{\cal C}_{NG}&=& \prod_{\gamma^m_{ G} \in \Re  \gamma^m} \, \gamma^m_{ G}\, K \,
 I_{x^6 x^8...x^d}\,,\nonumber\\
{\cal T}_{NG}&=& \gamma^0_{G} \prod_{\gamma^m_{G} \in \Im
 \gamma^m} \, K \, I_{x^0} I_{x^5 x^7...x^{d-1}}\,,\nonumber\\
{\cal P}^{(d-1)}_{NG} &=& \gamma^0_{G} \, \prod_{s=5}^{d}\, \gamma^s_{ G} I_{\vec{x}}\,,
\nonumber\\
{\bf \mathbb{C}}_{NG} &=&\prod_{\gamma^s_{ G} \in \Re  \gamma^s} \,\gamma^s_{ G}\,, 
 I_{x^6 x^8...x^d}\,,\quad
{\bf \mathbb{C}}_{NG} {\cal P}^{(d-1)}_{NG} = \gamma^0_{G} \, \prod_{
\gamma^s_{ G} \in \Im  \gamma^s, s=5}^{d}\, \gamma^s_{ G}\, I_{\vec{x}_{3}}\,
 I_{x^6 x^8...x^d}\,,
\nonumber\\
{\bf \mathbb{C}}_{NG} {\cal T}_{NG} {\cal P}^{(d-1)}_{NG} &=& \prod_{\gamma^s_{ G} \in 
\Im  \gamma^a} \,\gamma^a_{ G}\,I_x K\,.
\end{eqnarray}
%


%
\subsection{Representations in Grassmann and in Clifford space in $d=(13+1)$}
\label{so13+1}
In the {\it spin-charge-family} theory the starting dimension of space must be $\ge(13+1)$, in 
order that the theory manifests in $d=(3+1)$ all the observed properties of quarks and leptons,
 gauge and scalar fields (explaining the appearance of  higgs and 
the Yukawa couplings), offering as well the explanations for the observations in 
cosmology.

Let us therefore comment properties of representations in both spaces when $d=(13+1)$,
 if we analyze one group of "family members" of one of families in Clifford space, and 
one of the two representations of  $\frac{1}{2}\, \frac{d!}{\frac{d}{2}! \frac{d}{2}!}$.

{\bf a.} $\;\;$  Let us start 
with Clifford space~\cite{IARD2016,normaJMP2015,n2014matterantimatter,JMP2013,pikanorma,%
portoroz03,norma93}. Each "family" representation has $2^{\frac{d}{2} - 1}= 64$
 "family members". If we analyze this representation with respect to the subgroups $SO(3,1)$, 
$(SU(2)\times SU(2))$ of $SO(4)$ and ($SU(3)\times$ $U(1)$) of $SO(6)$ of  the 
Lorentz group $SO(13,1)$, we find that the representations have quantum numbers of all the so 
far observed quarks and leptons and antiquarks and antileptons, all with spin up and spin down, 
as well as of the left and right handedness, with the right handed neutrino included as the
member of this representation.

Let us make a choice of the "family", which follows by the application of $\tilde{S}^{15}$ on the 
"family", for which the creation operator of the right-handed neutrino with spin $\frac{1}{2}$ 
would be $ \stackrel{03}{(+i)}\,\stackrel{12}{(+)}|\stackrel{56}{(+)}\,
\stackrel{78}{(+)}||\stackrel{9 \;10}{(+)}\;\;\stackrel{11\;12}{(+)}\;\;\stackrel{13\;14}{(+)}$.
(The corresponding annihilation operator of this creation operator is  $\stackrel{13 \;14}{(-)}\;\;
\stackrel{11\;12}{(-)}\;\;\stackrel{9\;10}{(-)}||\stackrel{78}{(-)}\,\stackrel{56}{(-)}| 
\stackrel{12}{(-)}\,\stackrel{03}{(-i)}$). In Table~\ref{Table so13+1.} presented creation
operators for all the "family members" of this family follow 
by the application of $S^{ab}$ on 
$\tilde{S}^{15}$ $ \stackrel{03}{(+i)}\,\stackrel{12}{(+)}|\stackrel{56}{(+)}\,
\stackrel{78}{(+)}||\stackrel{9 \;10}{(+)}\;\;\stackrel{11\;12}{(+)}\;\;\stackrel{13\;14}{(+)}$.
(The annihilation operator of $\tilde{S}^{15}$ $ \stackrel{03}{(+i)}\,
\stackrel{12}{(+)}|\stackrel{56}{(+)}\,\stackrel{78}{(+)}||\stackrel{9\;10}{(+)}\;\;
\stackrel{11\;12}{(+)}\;\;\stackrel{13\;14}{(+)} $ is $\stackrel{13 \;14}{[-]}\;\;
\stackrel{11\;12}{[-]}\;\;\stackrel{9\;10}{(-)}||\stackrel{78}{(-)}\,\stackrel{56}{[+]}| 
\stackrel{12}{[+]}\,\stackrel{03}{(-i)}$.) 

This is the representation of 
Table~\ref{Table so13+1.}, in which all the 'family members'' of one "family"  are classified with 
respect to the subgroups $SO(3,1)\times SU(2) \times SU(2)\times SU(3) \times U(1)$.
The vacuum state on which the creation operators, represented in the third column, apply 
is defined in Eq.~(\ref{vac1}). All the creation operators of all the states are of an odd Clifford 
character, fulfilling together with the annihilation operators (which have as well the equivalent
odd Clifford character, since the Hermitian conjugation do not change the 
Clifford character) the requirements of Eq.~(\ref{ijthetaprod}). Since the Clifford even operators 
$S^{ab}$ and $\tilde{S}^{ab}$ do not change the Clifford character, all the creation and 
annihilation operators, obtained by products of $S^{ab}$ or $\tilde{S}^{ab}$ or both,
fulfill the requirements of Eq.~(\ref{ijthetaprod}).

We recognize in Table~\ref{Table so13+1.} that quarks distinguish from leptons only in the 
$SO(6)$ part of the creation operators. Quarks belong to the colour ($SU(3)$) triplet carrying
the "fermion" $(U(1))$ quantum number $\tau^{4} =\frac{1}{6}$, antiquarks belong to the colour 
antitriplet, carrying the "fermion" quantum number $\tau^{4} = -\frac{1}{6}$. Leptons belong 
to the colour ($SU(3)$) singlet, carrying the "fermion" $(U(1))$ quantum number $\tau^{4} =
 -\frac{1}{2}$, while antileptons belong to the colour antisinglet, carrying the "fermion" quantum 
number $\tau^{4} = \frac{1}{2}$. 
 
Let us also comment that the oddness and evenness of part of states in the subgroups of the 
$SO(13,1)$ group change: While quarks and leptons have in the part of $SO(6)$ an odd Clifford 
character, have antiquarks and antileptons in this part an even odd Clifford character. 
Correspondingly the Clifford character changes in the rest of subgroups 

\bottomcaption{\label{Table so13+1.}%
\tiny{
The left handed ($\Gamma^{(13,1)} = -1$~\cite{IARD2016})
multiplet of spinors --- the members of the fundamental representation of the $SO(13,1)$ group,
manifesting the subgroup $SO(7,1)$
 of the colour charged quarks and antiquarks and the colourless
leptons and antileptons --- is presented in the massless basis using the technique presented in
Refs.~\cite{nh02,nh03,IARD2016,normaJMP2015}.
It contains the left handed  ($\Gamma^{(3,1)}=-1$) 
 weak ($SU(2)_{I}$) charged  ($\tau^{13}=\pm \frac{1}{2}$, Eq.~(\ref{so42})),
and $SU(2)_{II}$ chargeless ($\tau^{23}=0$, Eq.~(\ref{so42}))
quarks and leptons and the right handed  ($\Gamma^{(3,1)}=1$) 
 weak  ($SU(2)_{I}$) chargeless and $SU(2)_{II}$
charged ($\tau^{23}=\pm \frac{1}{2}$) quarks and leptons, both with the spin $ S^{12}$  up and
down ($\pm \frac{1}{2}$, respectively). 
Quarks distinguish from leptons only in the $SU(3) \times U(1)$ part: Quarks are triplets
of three colours  ($c^i$ $= (\tau^{33}, \tau^{38})$ $ = [(\frac{1}{2},\frac{1}{2\sqrt{3}}),
(-\frac{1}{2},\frac{1}{2\sqrt{3}}), (0,-\frac{1}{\sqrt{3}}) $], Eq.~(\ref{so64}))
carrying  the "fermion charge" ($\tau^{4}=\frac{1}{6}$, Eq.~(\ref{so64})).
The colourless leptons carry the "fermion charge" ($\tau^{4}=-\frac{1}{2}$).
The same multiplet contains also the left handed weak ($SU(2)_{I}$) chargeless and $SU(2)_{II}$
charged antiquarks and antileptons and the right handed weak ($SU(2)_{I}$) charged and
$SU(2)_{II}$ chargeless antiquarks and antileptons.
Antiquarks distinguish from antileptons again only in the $SU(3) \times U(1)$ part: Antiquarks are
antitriplets, 
 carrying  the "fermion charge" ($\tau^{4}=-\frac{1}{6}$).
The anticolourless antileptons carry the "fermion charge" ($\tau^{4}=\frac{1}{2}$).
 $Y=(\tau^{23} + \tau^{4})$ is the hyper charge, the electromagnetic charge
is $Q=(\tau^{13} + Y$).
%
The vacuum state,
on which the nilpotents and projectors operate, is presented in Eq.~(\ref{vac1}).
The reader can find this  Weyl representation also in
Refs.~\cite{n2014matterantimatter,pikanorma,portoroz03,normaJMP2015} and the references
therein. }
}
\tablehead{\hline
i&$$&$|^a\psi_i>$&$\Gamma^{(3,1)}$&$ S^{12}$&
$\tau^{13}$&$\tau^{23}$&$\tau^{33}$&$\tau^{38}$&$\tau^{4}$&$Y$&$Q$\\
\hline
&& ${\rm (Anti)octet},\,\Gamma^{(7,1)} = (-1)\,1\,, \,\Gamma^{(6)} = (1)\,-1$&&&&&&&&& \\
&& ${\rm of \;(anti) quarks \;and \;(anti)leptons}$&&&&&&&&&\\
\hline\hline}
\tabletail{\hline \multicolumn{12}{r}{\emph{Continued on next page}}\\}
\tablelasttail{\hline}
\begin{center}
\tiny{
\begin{supertabular}{|r|c||c||c|c||c|c||c|c|c||r|r|}
1&$ u_{R}^{c1}$&$ \stackrel{03}{(+i)}\,\stackrel{12}{[+]}|
\stackrel{56}{[+]}\,\stackrel{78}{(+)}
||\stackrel{9 \;10}{(+)}\;\;\stackrel{11\;12}{[-]}\;\;\stackrel{13\;14}{[-]} $ &1&$\frac{1}{2}$&0&
$\frac{1}{2}$&$\frac{1}{2}$&$\frac{1}{2\,\sqrt{3}}$&$\frac{1}{6}$&$\frac{2}{3}$&$\frac{2}{3}$\\
\hline
2&$u_{R}^{c1}$&$\stackrel{03}{[-i]}\,\stackrel{12}{(-)}|\stackrel{56}{[+]}\,\stackrel{78}{(+)}
||\stackrel{9 \;10}{(+)}\;\;\stackrel{11\;12}{[-]}\;\;\stackrel{13\;14}{[-]}$&1&$-\frac{1}{2}$&0&
$\frac{1}{2}$&$\frac{1}{2}$&$\frac{1}{2\,\sqrt{3}}$&$\frac{1}{6}$&$\frac{2}{3}$&$\frac{2}{3}$\\
\hline
3&$d_{R}^{c1}$&$\stackrel{03}{(+i)}\,\stackrel{12}{[+]}|\stackrel{56}{(-)}\,\stackrel{78}{[-]}
||\stackrel{9 \;10}{(+)}\;\;\stackrel{11\;12}{[-]}\;\;\stackrel{13\;14}{[-]}$&1&$\frac{1}{2}$&0&
$-\frac{1}{2}$&$\frac{1}{2}$&$\frac{1}{2\,\sqrt{3}}$&$\frac{1}{6}$&$-\frac{1}{3}$&$-\frac{1}{3}$\\
\hline
4&$ d_{R}^{c1} $&$\stackrel{03}{[-i]}\,\stackrel{12}{(-)}|
\stackrel{56}{(-)}\,\stackrel{78}{[-]}
||\stackrel{9 \;10}{(+)}\;\;\stackrel{11\;12}{[-]}\;\;\stackrel{13\;14}{[-]} $&1&$-\frac{1}{2}$&0&
$-\frac{1}{2}$&$\frac{1}{2}$&$\frac{1}{2\,\sqrt{3}}$&$\frac{1}{6}$&$-\frac{1}{3}$&$-\frac{1}{3}$\\
\hline
5&$d_{L}^{c1}$&$\stackrel{03}{[-i]}\,\stackrel{12}{[+]}|\stackrel{56}{(-)}\,\stackrel{78}{(+)}
||\stackrel{9 \;10}{(+)}\;\;\stackrel{11\;12}{[-]}\;\;\stackrel{13\;14}{[-]}$&-1&$\frac{1}{2}$&
$-\frac{1}{2}$&0&$\frac{1}{2}$&$\frac{1}{2\,\sqrt{3}}$&$\frac{1}{6}$&$\frac{1}{6}$&$-\frac{1}{3}$\\
\hline
6&$d_{L}^{c1} $&$ - \stackrel{03}{(+i)}\,\stackrel{12}{(-)}|\stackrel{56}{(-)}\,\stackrel{78}{(+)}
||\stackrel{9 \;10}{(+)}\;\;\stackrel{11\;12}{[-]}\;\;\stackrel{13\;14}{[-]} $&-1&$-\frac{1}{2}$&
$-\frac{1}{2}$&0&$\frac{1}{2}$&$\frac{1}{2\,\sqrt{3}}$&$\frac{1}{6}$&$\frac{1}{6}$&$-\frac{1}{3}$\\
\hline
7&$ u_{L}^{c1}$&$ - \stackrel{03}{[-i]}\,\stackrel{12}{[+]}|\stackrel{56}{[+]}\,\stackrel{78}{[-]}
||\stackrel{9 \;10}{(+)}\;\;\stackrel{11\;12}{[-]}\;\;\stackrel{13\;14}{[-]}$ &-1&$\frac{1}{2}$&
$\frac{1}{2}$&0 &$\frac{1}{2}$&$\frac{1}{2\,\sqrt{3}}$&$\frac{1}{6}$&$\frac{1}{6}$&$\frac{2}{3}$\\
\hline
8&$u_{L}^{c1}$&$\stackrel{03}{(+i)}\,\stackrel{12}{(-)}|\stackrel{56}{[+]}\,\stackrel{78}{[-]}
||\stackrel{9 \;10}{(+)}\;\;\stackrel{11\;12}{[-]}\;\;\stackrel{13\;14}{[-]}$&-1&$-\frac{1}{2}$&
$\frac{1}{2}$&0&$\frac{1}{2}$&$\frac{1}{2\,\sqrt{3}}$&$\frac{1}{6}$&$\frac{1}{6}$&$\frac{2}{3}$\\
\hline\hline
\shrinkheight{0.3\textheight}
9&$ u_{R}^{c2}$&$ \stackrel{03}{(+i)}\,\stackrel{12}{[+]}|
\stackrel{56}{[+]}\,\stackrel{78}{(+)}
||\stackrel{9 \;10}{[-]}\;\;\stackrel{11\;12}{(+)}\;\;\stackrel{13\;14}{[-]} $ &1&$\frac{1}{2}$&0&
$\frac{1}{2}$&$-\frac{1}{2}$&$\frac{1}{2\,\sqrt{3}}$&$\frac{1}{6}$&$\frac{2}{3}$&$\frac{2}{3}$\\
\hline
10&$u_{R}^{c2}$&$\stackrel{03}{[-i]}\,\stackrel{12}{(-)}|\stackrel{56}{[+]}\,\stackrel{78}{(+)}
||\stackrel{9 \;10}{[-]}\;\;\stackrel{11\;12}{(+)}\;\;\stackrel{13\;14}{[-]}$&1&$-\frac{1}{2}$&0&
$\frac{1}{2}$&$-\frac{1}{2}$&$\frac{1}{2\,\sqrt{3}}$&$\frac{1}{6}$&$\frac{2}{3}$&$\frac{2}{3}$\\
\hline
11&$d_{R}^{c2}$&$\stackrel{03}{(+i)}\,\stackrel{12}{[+]}|\stackrel{56}{(-)}\,\stackrel{78}{[-]}
||\stackrel{9 \;10}{[-]}\;\;\stackrel{11\;12}{(+)}\;\;\stackrel{13\;14}{[-]}$
&1&$\frac{1}{2}$&0&
$-\frac{1}{2}$&$ - \frac{1}{2}$&$\frac{1}{2\,\sqrt{3}}$&$\frac{1}{6}$&$-\frac{1}{3}$&$-\frac{1}{3}$\\
\hline
12&$ d_{R}^{c2} $&$\stackrel{03}{[-i]}\,\stackrel{12}{(-)}|
\stackrel{56}{(-)}\,\stackrel{78}{[-]}
||\stackrel{9 \;10}{[-]}\;\;\stackrel{11\;12}{(+)}\;\;\stackrel{13\;14}{[-]} $
&1&$-\frac{1}{2}$&0&
$-\frac{1}{2}$&$-\frac{1}{2}$&$\frac{1}{2\,\sqrt{3}}$&$\frac{1}{6}$&$-\frac{1}{3}$&$-\frac{1}{3}$\\
\hline
13&$d_{L}^{c2}$&$\stackrel{03}{[-i]}\,\stackrel{12}{[+]}|\stackrel{56}{(-)}\,\stackrel{78}{(+)}
||\stackrel{9 \;10}{[-]}\;\;\stackrel{11\;12}{(+)}\;\;\stackrel{13\;14}{[-]}$
&-1&$\frac{1}{2}$&
$-\frac{1}{2}$&0&$-\frac{1}{2}$&$\frac{1}{2\,\sqrt{3}}$&$\frac{1}{6}$&$\frac{1}{6}$&$-\frac{1}{3}$\\
\hline
14&$d_{L}^{c2} $&$ - \stackrel{03}{(+i)}\,\stackrel{12}{(-)}|\stackrel{56}{(-)}\,\stackrel{78}{(+)}
||\stackrel{9 \;10}{[-]}\;\;\stackrel{11\;12}{(+)}\;\;\stackrel{13\;14}{[-]} $&-1&$-\frac{1}{2}$&
$-\frac{1}{2}$&0&$-\frac{1}{2}$&$\frac{1}{2\,\sqrt{3}}$&$\frac{1}{6}$&$\frac{1}{6}$&$-\frac{1}{3}$\\
\hline
15&$ u_{L}^{c2}$&$ - \stackrel{03}{[-i]}\,\stackrel{12}{[+]}|\stackrel{56}{[+]}\,\stackrel{78}{[-]}
||\stackrel{9 \;10}{[-]}\;\;\stackrel{11\;12}{(+)}\;\;\stackrel{13\;14}{[-]}$ &-1&$\frac{1}{2}$&
$\frac{1}{2}$&0 &$-\frac{1}{2}$&$\frac{1}{2\,\sqrt{3}}$&$\frac{1}{6}$&$\frac{1}{6}$&$\frac{2}{3}$\\
\hline
16&$u_{L}^{c2}$&$\stackrel{03}{(+i)}\,\stackrel{12}{(-)}|\stackrel{56}{[+]}\,\stackrel{78}{[-]}
||\stackrel{9 \;10}{[-]}\;\;\stackrel{11\;12}{(+)}\;\;\stackrel{13\;14}{[-]}$&-1&$-\frac{1}{2}$&
$\frac{1}{2}$&0&$-\frac{1}{2}$&$\frac{1}{2\,\sqrt{3}}$&$\frac{1}{6}$&$\frac{1}{6}$&$\frac{2}{3}$\\
\hline\hline
17&$ u_{R}^{c3}$&$ \stackrel{03}{(+i)}\,\stackrel{12}{[+]}|
\stackrel{56}{[+]}\,\stackrel{78}{(+)}
||\stackrel{9 \;10}{[-]}\;\;\stackrel{11\;12}{[-]}\;\;\stackrel{13\;14}{(+)} $ &1&$\frac{1}{2}$&0&
$\frac{1}{2}$&$0$&$-\frac{1}{\sqrt{3}}$&$\frac{1}{6}$&$\frac{2}{3}$&$\frac{2}{3}$\\
\hline
18&$u_{R}^{c3}$&$\stackrel{03}{[-i]}\,\stackrel{12}{(-)}|\stackrel{56}{[+]}\,\stackrel{78}{(+)}
||\stackrel{9 \;10}{[-]}\;\;\stackrel{11\;12}{[-]}\;\;\stackrel{13\;14}{(+)}$&1&$-\frac{1}{2}$&0&
$\frac{1}{2}$&$0$&$-\frac{1}{\sqrt{3}}$&$\frac{1}{6}$&$\frac{2}{3}$&$\frac{2}{3}$\\
\hline
19&$d_{R}^{c3}$&$\stackrel{03}{(+i)}\,\stackrel{12}{[+]}|\stackrel{56}{(-)}\,\stackrel{78}{[-]}
||\stackrel{9 \;10}{[-]}\;\;\stackrel{11\;12}{[-]}\;\;\stackrel{13\;14}{(+)}$&1&$\frac{1}{2}$&0&
$-\frac{1}{2}$&$0$&$-\frac{1}{\sqrt{3}}$&$\frac{1}{6}$&$-\frac{1}{3}$&$-\frac{1}{3}$\\
\hline
20&$ d_{R}^{c3} $&$\stackrel{03}{[-i]}\,\stackrel{12}{(-)}|
\stackrel{56}{(-)}\,\stackrel{78}{[-]}
||\stackrel{9 \;10}{[-]}\;\;\stackrel{11\;12}{[-]}\;\;\stackrel{13\;14}{(+)} $&1&$-\frac{1}{2}$&0&
$-\frac{1}{2}$&$0$&$-\frac{1}{\sqrt{3}}$&$\frac{1}{6}$&$-\frac{1}{3}$&$-\frac{1}{3}$\\
\hline
21&$d_{L}^{c3}$&$\stackrel{03}{[-i]}\,\stackrel{12}{[+]}|\stackrel{56}{(-)}\,\stackrel{78}{(+)}
||\stackrel{9 \;10}{[-]}\;\;\stackrel{11\;12}{[-]}\;\;\stackrel{13\;14}{(+)}$&-1&$\frac{1}{2}$&
$-\frac{1}{2}$&0&$0$&$-\frac{1}{\sqrt{3}}$&$\frac{1}{6}$&$\frac{1}{6}$&$-\frac{1}{3}$\\
\hline
22&$d_{L}^{c3} $&$ - \stackrel{03}{(+i)}\,\stackrel{12}{(-)}|\stackrel{56}{(-)}\,\stackrel{78}{(+)}
||\stackrel{9 \;10}{[-]}\;\;\stackrel{11\;12}{[-]}\;\;\stackrel{13\;14}{(+)} $&-1&$-\frac{1}{2}$&
$-\frac{1}{2}$&0&$0$&$-\frac{1}{\sqrt{3}}$&$\frac{1}{6}$&$\frac{1}{6}$&$-\frac{1}{3}$\\
\hline
23&$ u_{L}^{c3}$&$ - \stackrel{03}{[-i]}\,\stackrel{12}{[+]}|\stackrel{56}{[+]}\,\stackrel{78}{[-]}
||\stackrel{9 \;10}{[-]}\;\;\stackrel{11\;12}{[-]}\;\;\stackrel{13\;14}{(+)}$ &-1&$\frac{1}{2}$&
$\frac{1}{2}$&0 &$0$&$-\frac{1}{\sqrt{3}}$&$\frac{1}{6}$&$\frac{1}{6}$&$\frac{2}{3}$\\
\hline
24&$u_{L}^{c3}$&$\stackrel{03}{(+i)}\,\stackrel{12}{(-)}|\stackrel{56}{[+]}\,\stackrel{78}{[-]}
||\stackrel{9 \;10}{[-]}\;\;\stackrel{11\;12}{[-]}\;\;\stackrel{13\;14}{(+)}$&-1&$-\frac{1}{2}$&
$\frac{1}{2}$&0&$0$&$-\frac{1}{\sqrt{3}}$&$\frac{1}{6}$&$\frac{1}{6}$&$\frac{2}{3}$\\
\hline\hline
25&$ \nu_{R}$&$ \stackrel{03}{(+i)}\,\stackrel{12}{[+]}|
\stackrel{56}{[+]}\,\stackrel{78}{(+)}
||\stackrel{9 \;10}{(+)}\;\;\stackrel{11\;12}{(+)}\;\;\stackrel{13\;14}{(+)} $ &1&$\frac{1}{2}$&0&
$\frac{1}{2}$&$0$&$0$&$-\frac{1}{2}$&$0$&$0$\\
\hline
26&$\nu_{R}$&$\stackrel{03}{[-i]}\,\stackrel{12}{(-)}|\stackrel{56}{[+]}\,\stackrel{78}{(+)}
||\stackrel{9 \;10}{(+)}\;\;\stackrel{11\;12}{(+)}\;\;\stackrel{13\;14}{(+)}$&1&$-\frac{1}{2}$&0&
$\frac{1}{2}$ &$0$&$0$&$-\frac{1}{2}$&$0$&$0$\\
\hline
27&$e_{R}$&$\stackrel{03}{(+i)}\,\stackrel{12}{[+]}|\stackrel{56}{(-)}\,\stackrel{78}{[-]}
||\stackrel{9 \;10}{(+)}\;\;\stackrel{11\;12}{(+)}\;\;\stackrel{13\;14}{(+)}$&1&$\frac{1}{2}$&0&
$-\frac{1}{2}$&$0$&$0$&$-\frac{1}{2}$&$-1$&$-1$\\
\hline
28&$ e_{R} $&$\stackrel{03}{[-i]}\,\stackrel{12}{(-)}|
\stackrel{56}{(-)}\,\stackrel{78}{[-]}
||\stackrel{9 \;10}{(+)}\;\;\stackrel{11\;12}{(+)}\;\;\stackrel{13\;14}{(+)} $&1&$-\frac{1}{2}$&0&
$-\frac{1}{2}$&$0$&$0$&$-\frac{1}{2}$&$-1$&$-1$\\
\hline
29&$e_{L}$&$\stackrel{03}{[-i]}\,\stackrel{12}{[+]}|\stackrel{56}{(-)}\,\stackrel{78}{(+)}
||\stackrel{9 \;10}{(+)}\;\;\stackrel{11\;12}{(+)}\;\;\stackrel{13\;14}{(+)}$&-1&$\frac{1}{2}$&
$-\frac{1}{2}$&0&$0$&$0$&$-\frac{1}{2}$&$-\frac{1}{2}$&$-1$\\
\hline
30&$e_{L} $&$ - \stackrel{03}{(+i)}\,\stackrel{12}{(-)}|\stackrel{56}{(-)}\,\stackrel{78}{(+)}
||\stackrel{9 \;10}{(+)}\;\;\stackrel{11\;12}{(+)}\;\;\stackrel{13\;14}{(+)} $&-1&$-\frac{1}{2}$&
$-\frac{1}{2}$&0&$0$&$0$&$-\frac{1}{2}$&$-\frac{1}{2}$&$-1$\\
\hline
31&$ \nu_{L}$&$ - \stackrel{03}{[-i]}\,\stackrel{12}{[+]}|\stackrel{56}{[+]}\,\stackrel{78}{[-]}
||\stackrel{9 \;10}{(+)}\;\;\stackrel{11\;12}{(+)}\;\;\stackrel{13\;14}{(+)}$ &-1&$\frac{1}{2}$&
$\frac{1}{2}$&0 &$0$&$0$&$-\frac{1}{2}$&$-\frac{1}{2}$&$0$\\
\hline
32&$\nu_{L}$&$\stackrel{03}{(+i)}\,\stackrel{12}{(-)}|\stackrel{56}{[+]}\,\stackrel{78}{[-]}
||\stackrel{9 \;10}{(+)}\;\;\stackrel{11\;12}{(+)}\;\;\stackrel{13\;14}{(+)}$&-1&$-\frac{1}{2}$&
$\frac{1}{2}$&0&$0$&$0$&$-\frac{1}{2}$&$-\frac{1}{2}$&$0$\\
\hline\hline
33&$ \bar{d}_{L}^{\bar{c1}}$&$ \stackrel{03}{[-i]}\,\stackrel{12}{[+]}|
\stackrel{56}{[+]}\,\stackrel{78}{(+)}
||\stackrel{9 \;10}{[-]}\;\;\stackrel{11\;12}{(+)}\;\;\stackrel{13\;14}{(+)} $ &-1&$\frac{1}{2}$&0&
$\frac{1}{2}$&$-\frac{1}{2}$&$-\frac{1}{2\,\sqrt{3}}$&$-\frac{1}{6}$&$\frac{1}{3}$&$\frac{1}{3}$\\
\hline
34&$\bar{d}_{L}^{\bar{c1}}$&$\stackrel{03}{(+i)}\,\stackrel{12}{(-)}|\stackrel{56}{[+]}\,\stackrel{78}{(+)}
||\stackrel{9 \;10}{[-]}\;\;\stackrel{11\;12}{(+)}\;\;\stackrel{13\;14}{(+)}$&-1&$-\frac{1}{2}$&0&
$\frac{1}{2}$&$-\frac{1}{2}$&$-\frac{1}{2\,\sqrt{3}}$&$-\frac{1}{6}$&$\frac{1}{3}$&$\frac{1}{3}$\\
\hline
35&$\bar{u}_{L}^{\bar{c1}}$&$ - \stackrel{03}{[-i]}\,\stackrel{12}{[+]}|\stackrel{56}{(-)}\,\stackrel{78}{[-]}
||\stackrel{9 \;10}{[-]}\;\;\stackrel{11\;12}{(+)}\;\;\stackrel{13\;14}{(+)}$&-1&$\frac{1}{2}$&0&
$-\frac{1}{2}$&$-\frac{1}{2}$&$-\frac{1}{2\,\sqrt{3}}$&$-\frac{1}{6}$&$-\frac{2}{3}$&$-\frac{2}{3}$\\
\hline
36&$ \bar{u}_{L}^{\bar{c1}} $&$ - \stackrel{03}{(+i)}\,\stackrel{12}{(-)}|
\stackrel{56}{(-)}\,\stackrel{78}{[-]}
||\stackrel{9 \;10}{[-]}\;\;\stackrel{11\;12}{(+)}\;\;\stackrel{13\;14}{(+)} $&-1&$-\frac{1}{2}$&0&
$-\frac{1}{2}$&$-\frac{1}{2}$&$-\frac{1}{2\,\sqrt{3}}$&$-\frac{1}{6}$&$-\frac{2}{3}$&$-\frac{2}{3}$\\
\hline
37&$\bar{d}_{R}^{\bar{c1}}$&$\stackrel{03}{(+i)}\,\stackrel{12}{[+]}|\stackrel{56}{[+]}\,\stackrel{78}{[-]}
||\stackrel{9 \;10}{[-]}\;\;\stackrel{11\;12}{(+)}\;\;\stackrel{13\;14}{(+)}$&1&$\frac{1}{2}$&
$\frac{1}{2}$&0&$-\frac{1}{2}$&$-\frac{1}{2\,\sqrt{3}}$&$-\frac{1}{6}$&$-\frac{1}{6}$&$\frac{1}{3}$\\
\hline
38&$\bar{d}_{R}^{\bar{c1}} $&$ - \stackrel{03}{[-i]}\,\stackrel{12}{(-)}|\stackrel{56}{[+]}\,\stackrel{78}{[-]}
||\stackrel{9 \;10}{[-]}\;\;\stackrel{11\;12}{(+)}\;\;\stackrel{13\;14}{(+)} $&1&$-\frac{1}{2}$&
$\frac{1}{2}$&0&$-\frac{1}{2}$&$-\frac{1}{2\,\sqrt{3}}$&$-\frac{1}{6}$&$-\frac{1}{6}$&$\frac{1}{3}$\\
\hline
39&$ \bar{u}_{R}^{\bar{c1}}$&$\stackrel{03}{(+i)}\,\stackrel{12}{[+]}|\stackrel{56}{(-)}\,\stackrel{78}{(+)}
||\stackrel{9 \;10}{[-]}\;\;\stackrel{11\;12}{(+)}\;\;\stackrel{13\;14}{(+)}$ &1&$\frac{1}{2}$&
$-\frac{1}{2}$&0 &$-\frac{1}{2}$&$-\frac{1}{2\,\sqrt{3}}$&$-\frac{1}{6}$&$-\frac{1}{6}$&$-\frac{2}{3}$\\
\hline
40&$\bar{u}_{R}^{\bar{c1}}$&$\stackrel{03}{[-i]}\,\stackrel{12}{(-)}|\stackrel{56}{(-)}\,\stackrel{78}{(+)}
||\stackrel{9 \;10}{[-]}\;\;\stackrel{11\;12}{(+)}\;\;\stackrel{13\;14}{(+)}$
&1&$-\frac{1}{2}$&
$-\frac{1}{2}$&0&$-\frac{1}{2}$&$-\frac{1}{2\,\sqrt{3}}$&$-\frac{1}{6}$&$-\frac{1}{6}$&$-\frac{2}{3}$\\
\hline\hline
41&$ \bar{d}_{L}^{\bar{c2}}$&$ \stackrel{03}{[-i]}\,\stackrel{12}{[+]}|
\stackrel{56}{[+]}\,\stackrel{78}{(+)}
||\stackrel{9 \;10}{(+)}\;\;\stackrel{11\;12}{[-]}\;\;\stackrel{13\;14}{(+)} $
&-1&$\frac{1}{2}$&0&
$\frac{1}{2}$&$\frac{1}{2}$&$-\frac{1}{2\,\sqrt{3}}$&$-\frac{1}{6}$&$\frac{1}{3}$&$\frac{1}{3}$\\
\hline
42&$\bar{d}_{L}^{\bar{c2}}$&$\stackrel{03}{(+i)}\,\stackrel{12}{(-)}|\stackrel{56}{[+]}\,\stackrel{78}{(+)}
||\stackrel{9 \;10}{(+)}\;\;\stackrel{11\;12}{[-]}\;\;\stackrel{13\;14}{(+)}$
&-1&$-\frac{1}{2}$&0&
$\frac{1}{2}$&$\frac{1}{2}$&$-\frac{1}{2\,\sqrt{3}}$&$-\frac{1}{6}$&$\frac{1}{3}$&$\frac{1}{3}$\\
\hline
43&$\bar{u}_{L}^{\bar{c2}}$&$ - \stackrel{03}{[-i]}\,\stackrel{12}{[+]}|\stackrel{56}{(-)}\,\stackrel{78}{[-]}
||\stackrel{9 \;10}{(+)}\;\;\stackrel{11\;12}{[-]}\;\;\stackrel{13\;14}{(+)}$
&-1&$\frac{1}{2}$&0&
$-\frac{1}{2}$&$\frac{1}{2}$&$-\frac{1}{2\,\sqrt{3}}$&$-\frac{1}{6}$&$-\frac{2}{3}$&$-\frac{2}{3}$\\
\hline
44&$ \bar{u}_{L}^{\bar{c2}} $&$ - \stackrel{03}{(+i)}\,\stackrel{12}{(-)}|
\stackrel{56}{(-)}\,\stackrel{78}{[-]}
||\stackrel{9 \;10}{(+)}\;\;\stackrel{11\;12}{[-]}\;\;\stackrel{13\;14}{(+)} $
&-1&$-\frac{1}{2}$&0&
$-\frac{1}{2}$&$\frac{1}{2}$&$-\frac{1}{2\,\sqrt{3}}$&$-\frac{1}{6}$&$-\frac{2}{3}$&$-\frac{2}{3}$\\
\hline
45&$\bar{d}_{R}^{\bar{c2}}$&$\stackrel{03}{(+i)}\,\stackrel{12}{[+]}|\stackrel{56}{[+]}\,\stackrel{78}{[-]}
||\stackrel{9 \;10}{(+)}\;\;\stackrel{11\;12}{[-]}\;\;\stackrel{13\;14}{(+)}$
&1&$\frac{1}{2}$&
$\frac{1}{2}$&0&$\frac{1}{2}$&$-\frac{1}{2\,\sqrt{3}}$&$-\frac{1}{6}$&$-\frac{1}{6}$&$\frac{1}{3}$\\
\hline
46&$\bar{d}_{R}^{\bar{c2}} $&$ - \stackrel{03}{[-i]}\,\stackrel{12}{(-)}|\stackrel{56}{[+]}\,\stackrel{78}{[-]}
||\stackrel{9 \;10}{(+)}\;\;\stackrel{11\;12}{[-]}\;\;\stackrel{13\;14}{(+)} $
&1&$-\frac{1}{2}$&
$\frac{1}{2}$&0&$\frac{1}{2}$&$-\frac{1}{2\,\sqrt{3}}$&$-\frac{1}{6}$&$-\frac{1}{6}$&$\frac{1}{3}$\\
\hline
47&$ \bar{u}_{R}^{\bar{c2}}$&$\stackrel{03}{(+i)}\,\stackrel{12}{[+]}|\stackrel{56}{(-)}\,\stackrel{78}{(+)}
||\stackrel{9 \;10}{(+)}\;\;\stackrel{11\;12}{[-]}\;\;\stackrel{13\;14}{(+)}$
 &1&$\frac{1}{2}$&
$-\frac{1}{2}$&0 &$\frac{1}{2}$&$-\frac{1}{2\,\sqrt{3}}$&$-\frac{1}{6}$&$-\frac{1}{6}$&$-\frac{2}{3}$\\
\hline
48&$\bar{u}_{R}^{\bar{c2}}$&$\stackrel{03}{[-i]}\,\stackrel{12}{(-)}|\stackrel{56}{(-)}\,\stackrel{78}{(+)}
||\stackrel{9 \;10}{(+)}\;\;\stackrel{11\;12}{[-]}\;\;\stackrel{13\;14}{(+)}$
&1&$-\frac{1}{2}$&
$-\frac{1}{2}$&0&$\frac{1}{2}$&$-\frac{1}{2\,\sqrt{3}}$&$-\frac{1}{6}$&$-\frac{1}{6}$&$-\frac{2}{3}$\\
\hline\hline
49&$ \bar{d}_{L}^{\bar{c3}}$&$ \stackrel{03}{[-i]}\,\stackrel{12}{[+]}|
\stackrel{56}{[+]}\,\stackrel{78}{(+)}
||\stackrel{9 \;10}{(+)}\;\;\stackrel{11\;12}{(+)}\;\;\stackrel{13\;14}{[-]} $ &-1&$\frac{1}{2}$&0&
$\frac{1}{2}$&$0$&$\frac{1}{\sqrt{3}}$&$-\frac{1}{6}$&$\frac{1}{3}$&$\frac{1}{3}$\\
\hline
50&$\bar{d}_{L}^{\bar{c3}}$&$\stackrel{03}{(+i)}\,\stackrel{12}{(-)}|\stackrel{56}{[+]}\,\stackrel{78}{(+)}
||\stackrel{9 \;10}{(+)}\;\;\stackrel{11\;12}{(+)}\;\;\stackrel{13\;14}{[-]} $&-1&$-\frac{1}{2}$&0&
$\frac{1}{2}$&$0$&$\frac{1}{\sqrt{3}}$&$-\frac{1}{6}$&$\frac{1}{3}$&$\frac{1}{3}$\\
\hline
51&$\bar{u}_{L}^{\bar{c3}}$&$ - \stackrel{03}{[-i]}\,\stackrel{12}{[+]}|\stackrel{56}{(-)}\,\stackrel{78}{[-]}
||\stackrel{9 \;10}{(+)}\;\;\stackrel{11\;12}{(+)}\;\;\stackrel{13\;14}{[-]} $&-1&$\frac{1}{2}$&0&
$-\frac{1}{2}$&$0$&$\frac{1}{\sqrt{3}}$&$-\frac{1}{6}$&$-\frac{2}{3}$&$-\frac{2}{3}$\\
\hline
52&$ \bar{u}_{L}^{\bar{c3}} $&$ - \stackrel{03}{(+i)}\,\stackrel{12}{(-)}|
\stackrel{56}{(-)}\,\stackrel{78}{[-]}
||\stackrel{9 \;10}{(+)}\;\;\stackrel{11\;12}{(+)}\;\;\stackrel{13\;14}{[-]}  $&-1&$-\frac{1}{2}$&0&
$-\frac{1}{2}$&$0$&$\frac{1}{\sqrt{3}}$&$-\frac{1}{6}$&$-\frac{2}{3}$&$-\frac{2}{3}$\\
\hline
53&$\bar{d}_{R}^{\bar{c3}}$&$\stackrel{03}{(+i)}\,\stackrel{12}{[+]}|\stackrel{56}{[+]}\,\stackrel{78}{[-]}
||\stackrel{9 \;10}{(+)}\;\;\stackrel{11\;12}{(+)}\;\;\stackrel{13\;14}{[-]} $&1&$\frac{1}{2}$&
$\frac{1}{2}$&0&$0$&$\frac{1}{\sqrt{3}}$&$-\frac{1}{6}$&$-\frac{1}{6}$&$\frac{1}{3}$\\
\hline
54&$\bar{d}_{R}^{\bar{c3}} $&$ - \stackrel{03}{[-i]}\,\stackrel{12}{(-)}|\stackrel{56}{[+]}\,\stackrel{78}{[-]}
||\stackrel{9 \;10}{(+)}\;\;\stackrel{11\;12}{(+)}\;\;\stackrel{13\;14}{[-]} $&1&$-\frac{1}{2}$&
$\frac{1}{2}$&0&$0$&$\frac{1}{\sqrt{3}}$&$-\frac{1}{6}$&$-\frac{1}{6}$&$\frac{1}{3}$\\
\hline
55&$ \bar{u}_{R}^{\bar{c3}}$&$\stackrel{03}{(+i)}\,\stackrel{12}{[+]}|\stackrel{56}{(-)}\,\stackrel{78}{(+)}
||\stackrel{9 \;10}{(+)}\;\;\stackrel{11\;12}{(+)}\;\;\stackrel{13\;14}{[-]} $ &1&$\frac{1}{2}$&
$-\frac{1}{2}$&0 &$0$&$\frac{1}{\sqrt{3}}$&$-\frac{1}{6}$&$-\frac{1}{6}$&$-\frac{2}{3}$\\
\hline
56&$\bar{u}_{R}^{\bar{c3}}$&$\stackrel{03}{[-i]}\,\stackrel{12}{(-)}|\stackrel{56}{(-)}\,\stackrel{78}{(+)}
||\stackrel{9 \;10}{(+)}\;\;\stackrel{11\;12}{(+)}\;\;\stackrel{13\;14}{[-]} $&1&$-\frac{1}{2}$&
$-\frac{1}{2}$&0&$0$&$\frac{1}{\sqrt{3}}$&$-\frac{1}{6}$&$-\frac{1}{6}$&$-\frac{2}{3}$\\
\hline\hline
57&$ \bar{e}_{L}$&$ \stackrel{03}{[-i]}\,\stackrel{12}{[+]}|
\stackrel{56}{[+]}\,\stackrel{78}{(+)}
||\stackrel{9 \;10}{[-]}\;\;\stackrel{11\;12}{[-]}\;\;\stackrel{13\;14}{[-]} $ &-1&$\frac{1}{2}$&0&
$\frac{1}{2}$&$0$&$0$&$\frac{1}{2}$&$1$&$1$\\
\hline
58&$\bar{e}_{L}$&$\stackrel{03}{(+i)}\,\stackrel{12}{(-)}|\stackrel{56}{[+]}\,\stackrel{78}{(+)}
||\stackrel{9 \;10}{[-]}\;\;\stackrel{11\;12}{[-]}\;\;\stackrel{13\;14}{[-]}$&-1&$-\frac{1}{2}$&0&
$\frac{1}{2}$ &$0$&$0$&$\frac{1}{2}$&$1$&$1$\\
\hline
59&$\bar{\nu}_{L}$&$ - \stackrel{03}{[-i]}\,\stackrel{12}{[+]}|\stackrel{56}{(-)}\,\stackrel{78}{[-]}
||\stackrel{9 \;10}{[-]}\;\;\stackrel{11\;12}{[-]}\;\;\stackrel{13\;14}{[-]}$&-1&$\frac{1}{2}$&0&
$-\frac{1}{2}$&$0$&$0$&$\frac{1}{2}$&$0$&$0$\\
\hline
60&$ \bar{\nu}_{L} $&$ - \stackrel{03}{(+i)}\,\stackrel{12}{(-)}|
\stackrel{56}{(-)}\,\stackrel{78}{[-]}
||\stackrel{9 \;10}{[-]}\;\;\stackrel{11\;12}{[-]}\;\;\stackrel{13\;14}{[-]} $&-1&$-\frac{1}{2}$&0&
$-\frac{1}{2}$&$0$&$0$&$\frac{1}{2}$&$0$&$0$\\
\hline
61&$\bar{\nu}_{R}$&$\stackrel{03}{(+i)}\,\stackrel{12}{[+]}|\stackrel{56}{(-)}\,\stackrel{78}{(+)}
||\stackrel{9 \;10}{[-]}\;\;\stackrel{11\;12}{[-]}\;\;\stackrel{13\;14}{[-]}$&1&$\frac{1}{2}$&
$-\frac{1}{2}$&0&$0$&$0$&$\frac{1}{2}$&$\frac{1}{2}$&$0$\\
\hline
62&$\bar{\nu}_{R} $&$ - \stackrel{03}{[-i]}\,\stackrel{12}{(-)}|\stackrel{56}{(-)}\,\stackrel{78}{(+)}
||\stackrel{9 \;10}{[-]}\;\;\stackrel{11\;12}{[-]}\;\;\stackrel{13\;14}{[-]} $&1&$-\frac{1}{2}$&
$-\frac{1}{2}$&0&$0$&$0$&$\frac{1}{2}$&$\frac{1}{2}$&$0$\\
\hline
63&$ \bar{e}_{R}$&$\stackrel{03}{(+i)}\,\stackrel{12}{[+]}|\stackrel{56}{[+]}\,\stackrel{78}{[-]}
||\stackrel{9 \;10}{[-]}\;\;\stackrel{11\;12}{[-]}\;\;\stackrel{13\;14}{[-]}$ &1&$\frac{1}{2}$&
$\frac{1}{2}$&0 &$0$&$0$&$\frac{1}{2}$&$\frac{1}{2}$&$1$\\
\hline
64&$\bar{e}_{R}$&$\stackrel{03}{[-i]}\,\stackrel{12}{(-)}|\stackrel{56}{[+]}\,\stackrel{78}{[-]}
||\stackrel{9 \;10}{[-]}\;\;\stackrel{11\;12}{[-]}\;\;\stackrel{13\;14}{[-]}$&1&$-\frac{1}{2}$&
$\frac{1}{2}$&0&$0$&$0$&$\frac{1}{2}$&$\frac{1}{2}$&$1$\\
\hline
\end{supertabular}
}
\end{center}

Families are generated by $\tilde{S}^{ab}$ applying on any one of the  "family members".  
Again all the "family members" of this "family" follow by the application of all $S^{ab}$ (not 
belonging to Cartan subalgebra). 

The spontaneous break of symmetry from 
$SO(13,1)$ to $SO(7,1) \times SU(3) \times U(1)$, Refs.~\cite{IARD2016,n2014matterantimatter,%
normaJMP2015}, makes in the {\it spin-charge-family} theory  all the families, generated by  
$\tilde{S}^{mt}$ and $\tilde{S}^{st}$,
[$m=(0,1,2,3)$, $s=(5,6,7,8), t=(9,10,11,12,13,14)$], massive of the scale 
of $\ge 10^{16}$ GeV~\cite{DHN,DN012,%
familiesDNproc}. Correspondingly there are only eight families of
quarks and leptons, 
which split into two groups of four families, both manifesting  the symmetry $\widetilde{SU}(2)
\times \widetilde{SU}(2)$ $\times U(1)$. (The fourth of the lower four families is predicted to be 
observed at the LHC, the stable  of the upper four families
contributes to the dark  
matter~\cite{gn2009}.)

In the {\it spin-charge-family} theory fermions interact with only gravity, which manifests
after the break of the starting symmetry in $d=(3+1)$ as all the known vector gauge fields, 
ordinary gravity and the higgs and the Yukawa couplings~\cite{nd2017,IARD2016,%
n2014matterantimatter,normaJMP2015,n2012scalars}. There are scalar fields which bring masses 
to family members. The theory explains not only all the assumptions of the {\it standard model}
with the appearance of families, the vector gauge fields and the scalar fields, it also explains 
appearance of the dark matter~\cite{gn2009}, 
matter/antimatter asymmetry~\cite{n2014matterantimatter} and other phenomena, like the
miraculous cancellation of the triangle anomalies  in the {\it standard model}~\cite{nh2017}.

{\bf b.} $\;\;$ We compare representations of $SO(13,1)$ in Clifford space with those in 
Grassmann space. We have {\bf no "family" quantum numbers in Grassmann space}. 
We only have two groups of creation operators, defining --- when applied on the vacuum state
 $|1>$ --- $\frac{1}{2}$ $\frac{d!}{\frac{d}{2}! \frac{d}{2}!}$ equal in $d=(13+1)$ to $1716$
members in each of the two groups in comparison  in Clifford case with $64$ "family 
members" in one "family" and $64$ "families", which  the breaks of symmetry reduce to 
$8$ "families", making all the $(64 - 8)$ "families" massive and correspondingly not observable at low 
energies~(\cite{normaJMP2015,DHN} and the references therein).

Since the  $1716$ members are hard to be mastered, let us look therefore at each subgroup ---
$SU(3) \times U(1)$, $SO(3,1)$ and $SU(2)\times SU(2)$  of  $SO(13,1)$ --- separately.

Let us correspondingly analyze  the subgroups: $SO(6)$ from the point of view of
 the two subgroups $SU(3)\times U(1)$, and  $SO(7,1)$ from the point of view of
 the two subgroups $SO(3,1) \times SO(4)$, and let us also analyze $SO(4)$ as 
$SU(2)\times$ $SU(2)$.

\subsection{Examples of second quantizable states in Grassmann and in Clifford space} 
\label{examples}

We compare properties of representations in Grassmann and in Clifford space for several
choices of subgroups of  $SO(13,1)$ in the case 
that in both spaces creation and annihilation operators fulfill requirements of
 Eq.~(\ref{ijthetaprod}), that is that both kinds of states can be second quantized.
Let us again point out that in Grassmann case fermions carry integer spins, while in Clifford
case they carry half integer spin.

\subsubsection{ States in Grassmann and in Clifford space for $d=(5+1)$}
\label{examples51}

We study properties of representations of the subgroup $SO(5,1)$ (of the group $SO(13,1)$),
in Clifford and in Grassmann space, requiring that states can be in both spaces second quantized, 
fulfilling therefore Eq.~(\ref{ijthetaprod}).

{\bf a. }
In Clifford space there are $2^{\frac{d}{2}-1}$, each with  $2^{\frac{d}{2}-1}$ family 
members, that is $4$ families, each with $4$ members. All these sixteen states are of 
an odd Clifford character, since all can be obtained by products of $S^{ab}$, 
$\tilde{S}^{ab}$
or both from a Clifford odd starting state and are correspondingly second quantizable
as required in Eq.~(\ref{ijthetaprod}). All the states are  the eigenstates of the Cartan 
subalgebra of the Lorentz algebra in Clifford space, Eq.~(\ref{choicecartan}), solving the Weyl
equation for free massless spinors in Clifford space, Eq.~(\ref{Weyl}). 
 The four familes, with four members each, are presented in
Table~\ref{Table clifffourxfour51.}. All of these $16$ states are reachable from the first one
in each of the four families by $S^{ab}$, or by $\tilde{S}^{ab}$ if applied on any family member.

Each of these four families have positive and negative energy solutions, as presented in~%
\cite{nhds},  in Table I.. We present in Table~\ref{Table clifffourxfour51.} only states of a
positive energy, that is states  above the Dirac sea. The antiparticle states are reachable from
the particle states by the application of the operator $\mathbb{C}_{{\cal N}}\,
 $$ {\cal P}^{(d-1)}_{{\cal N}} = \gamma^0 \gamma^5 I_{\vec{x}_3} I_{x^6}$, keeping 
the spin $\frac{1}{2}$, while changing the charge from $\frac{1}{2}$ to $- \frac{1}{2}$. 
All the states above the Dirac sea are indeed the hole in the Dirac sea, as explained in
Ref.~\cite{nhds}.
\begin{table}
\begin{tiny}
  \centering
  \begin{tabular}{|c|c|c|c|c|c|c|c|}
    \hline\hline
    $$ & $\psi$ & $S^{03}$ & $S^{12}$ & $S^{56 }$ 
                         & $\Tilde{S}^{03}$ & $\Tilde{S}^{12}$ & $\Tilde{S}^{56 }$\\
    \hline\hline
        $\psi^{ I}_{1}$  & $\stackrel{03}{(+i)} \stackrel{12}{(+)} \stackrel{56}{(+)} $
                               & $\frac{i}{2}$ & $\frac{1}{2}$ &  $\frac{1}{2}$
    & $\frac{i}{2}$ & $\frac{1}{2}$ & $\frac{1}{2}$\\
        $\psi^{ I}_{2}$ &$\stackrel{03}{[-i]}\stackrel{12}{[-]} \stackrel{56}{(+)}$
        & $-\frac{i}{2}$ & $-\frac{1}{2}$ & $\frac{1}{2}$  & $\frac{i}{2}$ & $\frac{1}{2}$ & $\frac{1}{2}$                    \\
        $\psi^{ I}_{3}$ &$\stackrel{03}{[-i]} \stackrel{12}{(+)} \stackrel{56}{[-]} $
                               & $- \frac{i}{2}$ & $\frac{1}{2}$ & $-\frac{1}{2}$
      & $\frac{i}{2}$ & $\frac{1}{2}$ & $\frac{1}{2}$                              \\
        $\psi^{ I}_{4}$ &$\stackrel{03}{(+i)} \stackrel{12}{[-]} \stackrel{56}{[-]}$
      & $\frac{i}{2}$ & $-\frac{1}{2}$ & $-\frac{1}{2}$    & $\frac{i}{2}$ & $\frac{1}{2}$ & $\frac{1}{2}$                    \\
    \hline 
         $\psi^{II}_{1}$  & $\stackrel{03}{[+i]} \stackrel{12}{[+]} \stackrel{56}{(+)}$ 
                               & $\frac{i}{2}$ & $\frac{1}{2}$ &  $\frac{1}{2}$
      & $-\frac{i}{2}$ & $-\frac{1}{2}$ & $\frac{1}{2}$\\
        $\psi^{II}_{2}$  &$\stackrel{03}{(-i)} \stackrel{12}{(-)} \stackrel{56}{(+)}  $
       & $-\frac{i}{2}$ & $-\frac{1}{2}$ & $\frac{1}{2}$ & $-\frac{i}{2}$ & $-\frac{1}{2}$ & $\frac{1}{2}$                        \\
        $\psi^{II}_{3}$  & $\stackrel{03}{(-i)} \stackrel{12}{[+]} \stackrel{56}{[-]}$
         & $-\frac{i}{2}$ & $\frac{1}{2}$ & $-\frac{1}{2}$       & $-\frac{i}{2}$ & $-\frac{1}{2}$ & $\frac{1}{2}$               \\
        $\psi^{II}_{4}$  & $\stackrel{03}{[+i]}\stackrel{12}{(-)} \stackrel{56}{[-]}$
       & $\frac{i}{2}$ & $-\frac{1}{2}$ & $-\frac{1}{2}$         & $-\frac{i}{2}$ & $-\frac{1}{2}$ & $\frac{1}{2}$                       \\
       \hline
       $\psi^{III}_{1}$ & $\stackrel{03}{[+i]} \stackrel{12}{(+)} \stackrel{56}{[+]}$
                                 & $\frac{i}{2}$ & $\frac{1}{2}$ & $\frac{1}{2}$
       & $-\frac{i}{2}$ & $\frac{1}{2}$ & $-\frac{1}{2}$\\
        $\psi^{III}_{2}$ & $\stackrel{03}{(-i)} \stackrel{12}{[-]} \stackrel{56}{[+]}$
         & $-\frac{i}{2}$ & $-\frac{1}{2}$ & $\frac{1}{2}$  & $-\frac{i}{2}$ & $\frac{1}{2}$ & $-\frac{1}{2}$                              \\
       $\psi^{III}_{3}$ & $\stackrel{03}{(-i)} \stackrel{12}{(+)} \stackrel{56}{(-)} $
          & $-\frac{i}{2}$ & $\frac{1}{2}$ & $-\frac{1}{2}$   & $-\frac{i}{2}$ & $\frac{1}{2}$ & $-\frac{1}{2}$                   \\
       $\psi^{III}_{4}$ & $\stackrel{03}{[+i]} \stackrel{12}{[-]} \stackrel{56}{(-)} $
         & $\frac{i}{2}$ & $-\frac{1}{2}$ & $-\frac{1}{2}$ & $-\frac{i}{2}$ & $\frac{1}{2}$ & $-\frac{1}{2}$\\
        \hline
        $\psi^{IV}_{1}$ &  $\stackrel{03}{(+i)}\stackrel{12}{[+]} \stackrel{56}{[+]} $
                               & $\frac{i}{2}$ & $\frac{1}{2}$ & $\frac{1}{2}$ 
                              & $\frac{i}{2}$ & $-\frac{1}{2}$ & $-\frac{1}{2}$\\
        $\psi^{IV}_{2}$ & $\stackrel{03}{[-i]} \stackrel{12}{(-)}\stackrel{56}{[+]}  $ 
         & $-\frac{i}{2}$ & $-\frac{1}{2}$ & $\frac{1}{2}$ & $\frac{i}{2}$ & $-\frac{1}{2}$ & $-\frac{1}{2}$                     \\
       $\psi^{IV}_{3}$ & $\stackrel{03}{[-i]} \stackrel{12}{[+]}\stackrel{56}{(-)}  $
        & $-\frac{i}{2}$ & $\frac{1}{2}$ & $-\frac{1}{2}$  & $\frac{i}{2}$ & $-\frac{1}{2}$ & $-\frac{1}{2}$                              \\
       $\psi^{IV}_{4}$ & $\stackrel{03}{(+i)} \stackrel{12}{(-)} \stackrel{56}{(-)}  $ 
        & $\frac{i}{2}$ & $-\frac{1}{2}$ & $-\frac{1}{2}$  & $\frac{i}{2}$ & $-\frac{1}{2}$ & $-\frac{1}{2}$                       \\
    \hline\hline
  \end{tabular}
  \caption{\label{Table clifffourxfour51.} The four families, each with four members. For the 
choice $p^a=(p^0,0,0,p^3,0,0)$  have the first and the second member  the space part 
equal to $e^{- i |p^0| x^0+i |p^3|x^3}$ and $e^{- i |p^0| x^0-i |p^3|x^3}$,  
representing the particles with spin up and down, respectively. The third and the fourth member
represent the antiparticle states, with the space part equal to  $e^{- i |p^0| x^0-i |p^3|x^3}$
and $e^{- i |p^0| x^0+i |p^3|x^3}$, with the spin up and down respectively.
The antiparticle states follow from the particle state by the application of $\mathbb{C}_{{\cal N}}\,
 $$ {\cal P}^{(d-1)}_{{\cal N}} = \gamma^0 \gamma^5 I_{\vec{x}_3} I_{x^6}$.
The charge of the particle states is $\frac{1}{2}$, for antiparticle states $-\frac{1}{2}$.}
\end{tiny}
\end{table}


{\bf b.0}
In Grassmann space there are $ \frac{d!}{\frac{d}{2}! \frac{d}{2}!}$ second quantizable
states as required in Eq.~(\ref{ijthetaprod}), forming in $d=(5+1)$ two decuplets --- 
each with  $ \frac{1}{2} \,\frac{d!}{\frac{d}{2}! \frac{d}{2}!}$ states --- all are  the 
eigenstates of the Cartan subalgebra of the Lorentz algebra in (internal) Grassmann space. 
All the states of one (anyone of the two) decuplets are reachable by  the application of the 
operators ${\cal {\bf S}}^{ab}$ on a starting state. The two decouplets are presented in
Table~\ref{Table grassdecupletso51.}

Let us first find the solution of the equations of motion for free massless fermions,
 Eq.~(\ref{Weylgrass}),  with the momentum $p^a =(p^0, p^1,p^2,p^3,0,0)$. One obtains
for $\psi_I = \alpha (\theta^0-\theta^3) (\theta^1 +i \theta^2) (\theta^5 +i \theta^6)$
$+ \beta (\theta^0 \theta^3 + i\theta^1\theta^2) (\theta^5 +i \theta^6) +  $ 
$\gamma (\theta^0+\theta^3) (\theta^1 - i \theta^2) (\theta^5 +i \theta^6)$ the solution
\begin{eqnarray}
\label{5+1I}
\beta&=& \frac{2\gamma (p^1-i p^2)}{(p^0 -p^3)}=
\frac{2\gamma (p^0+ p^3)}{(p^1 +i p^2)}=- \frac{2\alpha (p^0- p^3)}{(p^1 -i p^2)}
= - \frac{2\alpha (p^1+i p^2)}{(p^0 +p^3)}\,,\nonumber\\
(p^0)^2 &=& (p^1)^2 + (p^2)^2 +(p^3)^2\,, \nonumber\\
\frac{\beta}{- \alpha } &=& \frac{2 (p^0- p^3)}{(p^1 -i p^2)}\,, \quad
\frac{\gamma}{- \alpha } =\frac{(p^0- p^3)^2}{(p^1 -i p^2)^2}\,.
\end{eqnarray}

One has for $p^0= |p^0|$ the positive energy solution, describing a fermion above the "Dirac sea", 
and for   $p^0=- |p^0|$ the negative energy solution, describing a fermion in  the "Dirac sea".
The "charge" of the "fermion" is $1$.  
Similarly one finds the solution for the other three states with the negative "charge" $-1$, again
with the positive and negative energy.
The space part of the "fermion" state is for "spin up" equal to 
$e^{- i |p^0| x^0+i \vec{p}\vec{x}}$, for his antiparticle for the same internal spin 
$e^{- i |p^0| x^0 - i \vec{p}\vec{x}}$.  

The discrete symmetry operator ${\bf \mathbb{C}}_{NG}$ ${\cal P}^{(d-1)}_{NG}$, which is 
in our case equal to $\gamma^0_{G} \gamma^5_{G}  I_{\vec{x}_{3}} I_{x^6}$, transforms
the  first state in Table~\ref{Table grassdecupletso51.} into the sixth, the  second state
into the fifth, the third state into the fourth, keeping the same spin while changing the "charge" of
the superposition of the three states $\psi_{Ip}$.
Both superposition of states, Eq.~(\ref{5+1I}) represent the positive energy states put on 
the top of the "Dirac" sea, the first 
describing a particle with "charge" $1$ and the second superposition of the second three states 
$\psi_{Ia}$, describing the antiparticle with the"charge" $-1$. 
We namely apply
 ${\underline {\bf \mathbb{C}}}_{{{\bf \cal NG}}}\, $$ {\cal P}^{(d-1)}_{{\cal NG}}$
 on ${\underline {\bf {\Huge \Psi}}}^{\dagger}_{p}[\Psi^{pos}_{I}]$ by applying 
 $\mathbb{C}_{{\cal NG}}\, $$ {\cal P}^{(d-1)}_{{\cal NG}}$ on $ \Psi^{pos}_{I}$ as follows: 
 ${\underline {\bf \mathbb{C}}}_{{{\bf \cal NG}}}\, $$ {\cal P}^{(d-1)}_{{\cal NG}}$ 
 ${\underline {\bf {\Huge \Psi}}}^{\dagger}_{p}[
 \Psi^{pos}_{I}]$ $({\underline {\bf \mathbb{C}}}_{{{\bf \cal NG}}}\, $
$ {\cal P}^{(d-1)}_{{\cal NG}})^{-1} =$ 
 ${\underline {\bf {\Huge \Psi}}}^{\dagger}_{aNG}$$[\mathbb{C}_{{\cal NG}}$
 ${\cal P}^{(d-1)}_{{\cal NG}}\Psi^{pos}_{1}]$.
 One recognizes that it is $\mathbb{C}_{{\cal NG}}\, $$ {\cal P}^{(d-1)}_{{\cal NG}}$ 
$ \Psi^{pos}_{I}=$
 $\Psi^{pos}_{II}$ (Table~\ref{Table grassdecupletso51.}), 
 which must be put on the top of the "Dirac" sea, representing the hole in the particular state 
 in the "Dirac" sea, which solves the corresponding  equation of motion for the negative energy.

 \begin{table}
 \begin{center}
\includegraphics{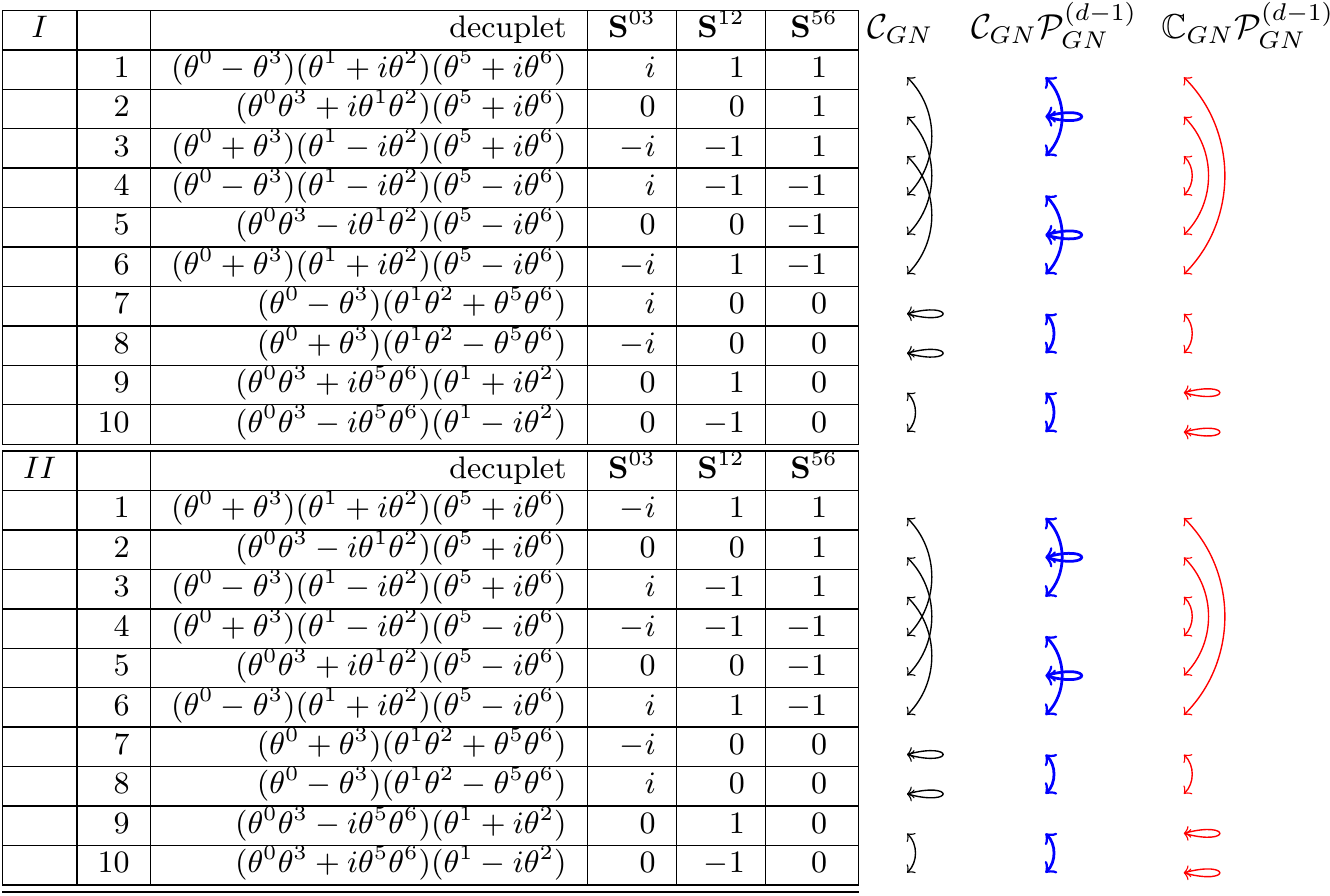}
 \end{center}
 \caption{\label{Table grassdecupletso51.}  The creation operators of the decuplet and the 
antidecuplet  of the orthogonal group $SO(5,1)$ in Grassmann space are presented. 
 Applying on the vacuum state $|\phi_{0}> = |1>$ the creation operators form eigenstates of
the Cartan subalgebra, Eq.~(\ref{choicecartan}), (${\cal {\bf S}}^{0 3}, {\cal {\bf S}}^{1 2}$, 
${\cal {\bf S}}^{5 6}$). The states within each decuplet are reachable from any 
member by ${\cal {\bf S}}^{ab}$. The product of the discrete operators
${\bf \mathbb{C}}_{NG}$ ($=\prod_{\Re \gamma^s} \, \gamma^s_{G}\, I_{x^6 x^8...x^d}$, 
denoted as ${\bf \mathbb{C}}$ in the last column) 
${\cal P}^{(d-1)}_{NG}$  ($ = \gamma^0_{G} \, \prod_{s=5}^{d}\, \gamma^s_{ G}
 I_{\vec{x}_{3}}$)
transforms, for example, $\psi^{I}_{1}$ into  $\psi^{I}_{6}$,  $\psi^{I}_{2}$  into  $\psi^{I}_{5}$ 
and $\psi^{I}_{3}$ into  $\psi^{I}_{4}$. Solutions of the Weyl equation, Eq.~(\ref{Weylgrass}), 
with the negative energies belong to the "Grassmann sea", with the positive energy to the particles 
and antiparticles.
Also the application of the discrete operators ${\cal C}_{GN}$, Eq.~(\ref{calCPTNG}) and 
${\cal C}_{NG}$ ${\cal P}^{(d-1)}_{NG}$, Eq.~(\ref{calCPTNG}) is demonstrated.
%
%
}
 \end{table}
\subsubsection{Properties of  $SO(6)$ in Grassmann and in Clifford space when $SO(6)$ is embedded
into $SO(13,1)$}
\label{so6}
{\bf a.} 
Let us first repeat properties of the $SO(6)$ part of the $SO(13,1)$ representation of $64$ 
"family members" in Clifford space, presented in Table~\ref{Table so13+1.}. As seen in
Table~\ref{Table so13+1.} there are one quadruplet ($2^{\frac{d}{2}-1}=4$) --- 
($\stackrel{9 \;10}{(+)}\;\;\stackrel{11\;12}{[-]}\;\;\stackrel{13\;14}{[-]} $,  
$\stackrel{9 \;10}{[-]}\;\;\stackrel{11\;12}{(+)}\;\;\stackrel{13\;14}{[-]}$,
$\stackrel{9 \;10}{[-]}\;\;\stackrel{11\;12}{[-]}\;\;\stackrel{13\;14}{(+)} $,
$\stackrel{9 \;10}{(+)}\;\;\stackrel{11\;12}{(+)}\;\;\stackrel{13\;14}{(+)}$), representing
quarks and leptons --- and one antiquadruplet --- 
($\stackrel{9 \;10}{[-]}\;\;\stackrel{11\;12}{(+)}\;\;\stackrel{13\;14}{(+)} $,  
$\stackrel{9 \;10}{(+)}\;\;\stackrel{11\;12}{[-]}\;\;\stackrel{13\;14}{(+)}$,
$\stackrel{9 \;10}{(+)}\;\;\stackrel{11\;12}{(+)}\;\;\stackrel{13\;14}{[-]} $,
$\stackrel{9 \;10}{[-]}\;\;\stackrel{11\;12}{[-]}\;\;\stackrel{13\;14}{[-]}$), representing
antiquarks and antileptons, which both belong to the $64^{th}$-plet, if $SO(6)$ is embedded into 
$SO(13,1)$. The creation operators (and correspondingly their annihilation operators) have 
for $32$ members (representing quarks and leptons) the $SO(6)$ part of an odd Clifford 
character (and can be correspondingly second quantized (by itselves~\cite{nh2018} or) together
with the rest of space, manifesting $SO(7,1)$ (since it has an even Clifford character).  The rest 
of $32$ creation operators 
(representing antiquarks and antileptons) has in the $SO(6)$ part an even Clifford character and 
correspondingly in the rest of the Clifford space in $SO(7,1)$ an odd Clifford character.

Let us discuss the case with the quadruplet of $SO(6)$ with an odd Clifford character. From the 
point of view 
of the subgroups $SU(3)$ (the colour subgroup) and $U(1)$ (the $U(1)$ subgroup carrying the
 "fermion"  quantum number),  the quadruplet consists of one $SU(3)$ singlet with the 
"fermion" quantum number $-\frac{1}{2}$ and one triplet  with the "fermion" quantum number 
$\frac{1}{6}$. The Clifford even $SO(7,1)$ part of $SO(13,1)$ define together with the Clifford 
odd  $SO(6)$ part the quantum numbers of the right handed 
quarks and  leptons and of the left handed quarks and leptons of the {\it standard model}, the
left handed weak charged and the right handed weak chargeless. 

 In the same representation of $SO(13,1)$ there is also one antiquadruplet, which has the even 
Clifford character of $SO(6)$ part and the odd Clifford character in the $SO(7,1)$ part of the 
$SO(13,1)$. The antiquadruplet of the $SO(6)$ part consists of one 
$SU(3)$ antisinglet with the "fermion" quantum number $\frac{1}{2}$ and one antitriplet 
 with the "fermion" quantum number $-\frac{1}{6}$. The $SO(7,1) \times SO(6)$ antiquadruplet
 of $SO(13,1)$ carries quantum numbers of left handed weak chargeless antiquarks and 
antileptons and of the right handed weak charged antiquarks and antileptons of the 
{\it standard model}. 

Both, quarks and leptons and antiquarks and antileptons, belong to the same representation of 
$SO(13,1)$,
explaining the miraculous cancellation of the triangle anomalies in the {\it standard model}
without connecting by hand the handedness and the charges of quarks and leptons~\cite{nh2017},
as it must be done in the $SO(10)$ models. 

{\bf b.}
In Grassmann space there are one ($\frac{1}{2}\,\frac{d!}{\frac{d}{2}! \frac{d}{2}!}= 10$)
decuplet representation of $SO(6)$ and one antidecuplet, both presented in 
Table~\ref{Table grassdecuplet.}. To be able to second quantize the theory,
the whole representation must be Grassmann odd. Both decuplets in 
Table~\ref{Table grassdecuplet.} have an odd Grassmann character, what means that products
of eigenstates of the Cartan subalgebra in the rest of Grassmann space must be of an 
Grassmann even character to be second quantizable. Both decuplets would, however, appear in 
the same representation of $SO(13,1)$, and one can expect also decuplets of an even
Grassmann character, if $SO(6)$ is embedded into $SO(13,1)$%
~\footnote{This can easily be understood, if we look at the subgroups of the group $SO(6)$. 
{\bf i.} $\,$ Let us look at the subgroup $SO(2)$. There are two creation operators of an odd 
Grassmann character, in this case $(\theta^{9} - i \theta^{10})$ and $(\theta^{9} +
 i \theta^{10})$. Both appear in either decuplet or in antidecuplet --- together with 
$\theta^{9} \theta^{10}$ with an even Grassmann character ---
multiplied by the part appearing from the rest of space $d=(11,12,13,14)$. But if $SO(2)$ is 
not embedded in $SO(6)$, then the two states, corresponding to the creation operators,  
$(\theta^{9} \mp i \theta^{10})$, belong to different representations, and so is  
$\theta^{9} \theta^{10}$. 
{\bf ii.} $\,$ Similarly we see, if we consider the subgroup $SO(4)$ of the group 
$SO(6)$. All six states, $(\theta^{9} + i \theta^{10})
\cdot (\theta^{11} + i \theta^{12})$, $(\theta^{9} - i \theta^{10})
\cdot (\theta^{11} - i \theta^{12})$, $(\theta^{9} \theta^{10} + \theta^{11} \theta^{12})$,
 $(\theta^{9} + i \theta^{10})\cdot (\theta^{11} - i \theta^{12})$, 
$(\theta^{9} - i \theta^{10})\cdot (\theta^{11} + i \theta^{12})$, $(\theta^{9} \theta^{10} - 
\theta^{11} \theta^{12})$, appear in the 
decuplet and in the antidecuplet, multiplied with the part appearing from the rest of space, in this 
case in $d=(13,14)$, if $SO(4)$ is embedded in $SO(6)$. But, in $d=4$ 
space there are two decoupled groups of three states~\cite{norma93}: 
[$(\theta^{9} + i  \theta^{10})\cdot (\theta^{11} + i  \theta^{12})$, $(\theta^{9} \theta^{10}
 + \theta^{11} \theta^{12})$, $(\theta^{9} - i  \theta^{10}) \cdot
 (\theta^{11} - i \theta^{12})$] and [$(\theta^{9} - i  \theta^{10}) \cdot (\theta^{11} + 
i  \theta^{12})$, $(\theta^{9} \theta^{10} - \theta^{11} \theta^{12})$, $(\theta^{9} + i  
\theta^{10}) \cdot (\theta^{11} - i \theta^{12})$]. Neither of these six members could be 
second quantized in $d=4$ alone.}.

With respect to $SU(3)\times U(1)$ subgroups of the group $SO(6)$  the decuplet manifests as one 
singlet, one triplet and one sextet, while the antidecuplet manifests as one antisinglet, one 
antitriplet and one antisextet. All the corresponding quantum numbers of either the Cartan 
subalgebra operators or of the corresponding diagonal operators  of the $SU(3)$ or $U(1)$ subgroups
are presented in Table~\ref{Table grassdecuplet.}. 
 \begin{table}
 \begin{center}
\begin{tiny}
 \begin{tabular}{|c|r|r|r|r|r|r|r|r|}
 \hline
$I$& &$\rm{decuplet}$&${\cal {\bf S}}^{9\,10}$&${\cal {\bf S}}^{11\,12}$&
${\cal {\bf S}}^{13\,14}$&${\bf \tau^{4}}$&${\bf \tau^{33}}$& ${\bf \tau^{38}}$\\
 \hline 
& $1$  & $ (\theta^{9} + i \theta^{10}) (\theta^{11} + i \theta^{12})
 (\theta^{13} + i \theta^{14})$ &$1$&$1$&$1$&$-1$&$0$&$0$\\
\hline
&$2$  & $ (\theta^{9} + i \theta^{10}) (\theta^{11}\theta^{12} +
 \theta^{13} \theta^{14})$ &$1$&$0$&$0$&$-\frac{1}{3}$&$+\frac{1}{2}$&$+
\frac{1}{2\sqrt{3}}$\\
\hline
&$3$  & $ (\theta^{9} + i \theta^{10}) (\theta^{11} - i \theta^{12})
 (\theta^{13} - i \theta^{14})$ &$1$&$-1$&$-1$&$ +\frac{1}{3}$&$+ 1$&$+
\frac{1}{\sqrt{3}}$\\
\hline
& $4$  & $ (\theta^{9} \theta^{10} + \theta^{11} \theta^{12})
 (\theta^{13} + i \theta^{14})$ &$0$&$0$&$1$&$-\frac{1}{3}$&$0$&$-\frac{1}{\sqrt{3}}$\\
\hline
&$5$  & $ (\theta^{9} - i \theta^{10}) (\theta^{11} - i \theta^{12})
 (\theta^{13} + i \theta^{14})$ &$-1$&$-1$&$-1$&$ +\frac{1}{3}$&$0$&$-\frac{2}{\sqrt{3}}$\\
\hline
&$6$  & $ (\theta^{11} + i \theta^{12}) (\theta^{9}\theta^{10} 
+ \theta^{13} \theta^{14})$ &$0$&$1$&$0$&$-\frac{1}{3}$&$-\frac{1}{2}$&$+\frac{1}{2\sqrt{3}}$\\
\hline
&$7$  & $ (\theta^{9} - i \theta^{10}) (\theta^{11} + i \theta^{12})
 (\theta^{13} - i \theta^{14})$ &$-1$&$1$&$-1$&$ +\frac{1}{3}$&$- 1$&$+\frac{1}{\sqrt{3}}$\\
\hline
&$8$  & $ (\theta^{9} \theta^{10} - \theta^{11} \theta^{12})
 (\theta^{13} - i \theta^{14})$ &$0$&$0$&$-1$&$+\frac{1}{3}$&$0$&$+\frac{1}{\sqrt{3}}$\\
\hline
& $9$  & $ (\theta^{9} \theta^{10} - \theta^{13} \theta^{14})
 (\theta^{11} - i \theta^{12})$ &$0$&$-1$&$0$&$+\frac{1}{3}$&$+\frac{1}{2}$&$-\frac{1}{2\sqrt{3}}$\\
\hline
&$10$  & $ (\theta^{9} - i \theta^{10}) (\theta^{11}\theta^{12} -
 \theta^{13} \theta^{14})$ &$-1$&$0$&$0$&$+\frac{1}{3}$&$-\frac{1}{2}$&$-\frac{1}{2\sqrt{3}}$\\
\hline \hline 
 $II$& &$\rm{decuplet}$&${\cal {\bf S}}^{9\,10}$&${\cal {\bf S}}^{11\,12}$&
${\cal {\bf S}}^{13\,14}$&${\bf \tau^{4}}$&${\bf \tau^{33}}$& ${\bf \tau^{38}}$\\
\hline
& $1$  & $ (\theta^{9} - i \theta^{10}) (\theta^{11} - i \theta^{12})
 (\theta^{13} - i \theta^{14})$ &$-1$&$-1$&$-1$&$+1$&$0$&$0$\\
\hline
&$2$  & $ (\theta^{9} - i \theta^{10}) (\theta^{11}\theta^{12} +
 \theta^{13} \theta^{14})$ &$-1$&$0$&$0$&$+\frac{1}{3}$&$-\frac{1}{2}$&$-\frac{1}{2\sqrt{3}}$\\
\hline
&$3$  & $ (\theta^{9} - i \theta^{10}) (\theta^{11} + i \theta^{12})
 (\theta^{13} + i \theta^{14})$ &$-1$&$1$&$1$&$ -\frac{1}{3}$&$- 1$&$-\frac{1}{\sqrt{3}}$\\
\hline
& $4$  & $ (\theta^{9} \theta^{10} + \theta^{11} \theta^{12})
 (\theta^{13} - i \theta^{14})$ &$0$&$0$&$-1$&$+\frac{1}{3}$&$0$&$+\frac{1}{\sqrt{3}}$\\
\hline
&$5$  & $ (\theta^{9} + i \theta^{10}) (\theta^{11} + i \theta^{12})
 (\theta^{13} - i \theta^{14})$ &$1$&$1$&$-1$&$ -\frac{1}{3}$&$0$&$+\frac{2}{\sqrt{3}}$\\
\hline
&$6$  & $ (\theta^{11} - i \theta^{12}) (\theta^{9}\theta^{10} +
 \theta^{13} \theta^{14})$ &$0$&$-1$&$0$&$+\frac{1}{3}$&$+\frac{1}{2}$&$-\frac{1}{2\sqrt{3}}$\\
\hline
&$7$  & $ (\theta^{9} + i \theta^{10}) (\theta^{11} - i \theta^{12})
 (\theta^{13} + i \theta^{14})$ &$1$&$-1$&$1$&$ -\frac{1}{3}$&$+ 1$&$-\frac{1}{\sqrt{3}}$\\
\hline
&$8$  & $ (\theta^{9} \theta^{10} - \theta^{11} \theta^{12})
 (\theta^{13} + i \theta^{14})$ &$0$&$0$&$1$&$-\frac{1}{3}$&$0$&$-\frac{1}{\sqrt{3}}$\\
\hline
& $9$  & $ (\theta^{9} \theta^{10} - \theta^{13} \theta^{14})
 (\theta^{11} + i \theta^{12})$ &$0$&$1$&$0$&$-\frac{1}{3}$&$-\frac{1}{2}$&$+\frac{1}{2\sqrt{3}}$\\
\hline
&$10$  & $ (\theta^{9} + i \theta^{10}) (\theta^{11}\theta^{12} -
 \theta^{13} \theta^{14})$ &$1$&$0$&$0$&$-\frac{1}{3}$&$+\frac{1}{2}$&$+\frac{1}{2\sqrt{3}}$\\
 \hline
 \end{tabular}
\end{tiny}
 \end{center}
 \caption{\label{Table grassdecuplet.}  The creation operators of the decuplet and the 
antidecuplet  of the orthogonal group $SO(6)$ in Grassmann space are presented.  Applying on the vacuum state $|\phi_{0}> = |1>$ the creation operators form eigenstates of the Cartan subalgebra, 
Eq.~(\ref{choicecartan}), (${\cal {\bf S}}^{9\,10}, {\cal {\bf S}}^{11\,12}$, 
${\cal {\bf S}}^{13\,14}$). The states within each decouplet are reachable from any 
member by ${\cal {\bf S}}^{ab}$. The quantum numbers (${\bf \tau^{33}}, {\bf \tau^{38}}$) and 
${\bf \tau^{4}}$ of 
the subgroups $SU(3)$ and  $U(1)$ of the group $SO(6)$ are also presented, Eq.~(\ref{so64}).
%
}
 \end{table}
%

While in Clifford case the representations of $SO(6)$, if  the group $SO(6)$ is embedded into  
$SO(13,1)$,  are defining a Clifford odd quadruplet and a Clifford even antiquadruplet, the 
representations in Grassmann case 
define one decuplet and one antidecuplet, both of the same Grassmann character, the odd one in 
our case. The two quadruplets in Clifford case manifest 
with respect to the subgroups $SU(3)$ and $U(1)$ as a triplet and a singlet, and as an antitriplet
and an antisinglet, respectively. In Grassmann case the two decuplets manifest with 
respect to the subgroups $SU(3)$ and $U(1)$ as a (triplet, singlet, sextet) and  as an (antitriplet,
antisinglet, antisextet), respectively. The corresponding multiplets are presented in 
Table~\ref{Table grasssextet.}. 
 \begin{table}
 \begin{center}
\begin{tiny}
 \begin{tabular}{|c|r|r|r|r|r|}
 \hline \hline
$I$& &&${\bf \tau^{4}}$&${\bf \tau^{33}}$& ${\bf \tau^{38}}$\\
 \hline 
$\rm{singlet}$&$ $  & $ (\theta^{9} + i \theta^{10}) (\theta^{11} + i \theta^{12})
 (\theta^{13} + i \theta^{14})$ &$-1$&$0$&$0$\\
\hline\hline
$\rm{triplet}$&$1$  & $ (\theta^{9} + i \theta^{10}) (\theta^{11}\theta^{12} +
 \theta^{13} \theta^{14}) $ &$-\frac{1}{3}$&$+\frac{1}{2}$&$+\frac{1}{2\sqrt{3}}$\\
\hline
&$2$  & $ (\theta^{9} \theta^{10} + \theta^{11} \theta^{12})
 (\theta^{13} + i \theta^{14}) $ &$-\frac{1}{3}$&$0$&$-\frac{1}{\sqrt{3}}$\\
\hline
&$3$  & $ (\theta^{11} + i \theta^{12}) (\theta^{9}\theta^{10} 
+ \theta^{13} \theta^{14}) $ &$-\frac{1}{3}$&$-\frac{1}{2}$&$+\frac{1}{2\sqrt{3}}$\\
\hline \hline
$\rm{sextet}$&$1$  & $(\theta^{9}+ i \theta^{10}) (\theta^{11} - i \theta^{12})  
(\theta^{13} - i \theta^{14})
$ &$\frac{1}{3}$&$+1$&$+\frac{1}{\sqrt{3}}$\\
\hline
&$2$  & $ (\theta^{9} - i \theta^{10}) (\theta^{11} - i \theta^{12})
 (\theta^{13} + i \theta^{14}) $ &$\frac{1}{3}$&$0$&$-\frac{2}{\sqrt{3}}$\\
\hline
&$3$  & $ (\theta^{9} - i \theta^{10}) (\theta^{11} + i \theta^{12})
 (\theta^{13} - i \theta^{14}) $ &$\frac{1}{3}$&$-1$&$+\frac{1}{\sqrt{3}}$\\
\hline
&$4$  & $ (\theta^{9} \theta^{10} - \theta^{11} \theta^{12})
 (\theta^{13} - i \theta^{14}) $ &$\frac{1}{3}$&$0$&$+\frac{1}{\sqrt{3}}$\\
\hline
&$5$  & $(\theta^{9} \theta^{10} - \theta^{13} \theta^{14})
 (\theta^{11} - i \theta^{12}) $ &$\frac{1}{3}$&$+\frac{1}{2}$&$-\frac{1}{2\sqrt{3}}$\\
\hline
&$6$  & $(\theta^{9} - i \theta^{10}) (\theta^{11}\theta^{12} -
 \theta^{13} \theta^{14}) $ &$\frac{1}{3}$&$-\frac{1}{2}$&$-\frac{1}{2\sqrt{3}}$\\
\hline\hline
$II$&&&${\bf \tau^{4}}$&${\bf \tau^{33}}$& ${\bf \tau^{38}}$\\
 \hline 
$\rm{antisinglet}$&$ $  & $ (\theta^{9} - i \theta^{10}) (\theta^{11} - i \theta^{12})
 (\theta^{13} - i \theta^{14})$ &$+1$&$0$&$0$\\
\hline
\hline
$\rm{antitriplet}$&$1$  & $ (\theta^{9} - i \theta^{10}) (\theta^{11}\theta^{12} +
 \theta^{13} \theta^{14})$ &$+\frac{1}{3} $ &$-\frac{1}{2}$&$-\frac{1}{2\sqrt{3}}$\\
\hline
&$2$  & $  (\theta^{9} \theta^{10} + \theta^{11} \theta^{12})
 (\theta^{13} - i \theta^{14})$ &$+\frac{1}{3}$&$0$&$+\frac{1}{\sqrt{3}}$\\
\hline
&$3$  & $(\theta^{11} - i \theta^{12}) (\theta^{9}\theta^{10} +
 \theta^{13} \theta^{14}) $ &$+\frac{1}{3}$&$+\frac{1}{2}$&$-\frac{1}{2\sqrt{3}}$\\
\hline \hline
$\rm{antisextet}$&$1$  & $(\theta^{9} - i \theta^{10}) (\theta^{11} + i \theta^{12})
 (\theta^{13} + i \theta^{14})$ &$ -\frac{1}{3} $ &$-1$&$-\frac{1}{\sqrt{3}}$\\
\hline
&$2$  & $(\theta^{9} + i \theta^{10}) (\theta^{11} + i \theta^{12})
 (\theta^{13} - i \theta^{14}) $ &$-\frac{1}{3}$&$0$&$+\frac{2}{\sqrt{3}}$\\
\hline
&$3$  & $ (\theta^{9} + i \theta^{10}) (\theta^{11} - i \theta^{12})
 (\theta^{13} + i \theta^{14}) $ &$-\frac{1}{3}$&$+1$&$-\frac{1}{\sqrt{3}}$\\
\hline
&$4$  & $ (\theta^{9} \theta^{10} - \theta^{11} \theta^{12})
 (\theta^{13} + i \theta^{14}) $ &$-\frac{1}{3}$&$0$&$-\frac{1}{\sqrt{3}}$\\
\hline
&$5$  & $ (\theta^{9} \theta^{10} - \theta^{13} \theta^{14})
 (\theta^{11} + i \theta^{12}) $ &$-\frac{1}{3}$&$-\frac{1}{2}$&$+\frac{1}{2\sqrt{3}}$\\
\hline
&$6$  & $ (\theta^{9} + i \theta^{10}) (\theta^{11}\theta^{12} -
 \theta^{13} \theta^{14}) $ &$-\frac{1}{3}$&$+\frac{1}{2}$&$+\frac{1}{2\sqrt{3}}$\\
\hline
 \end{tabular}
\end{tiny}
 \end{center}
 \caption{\label{Table grasssextet.} The creation operators in Grassmann space of the decuplet 
of Table~\ref{Table grassdecuplet.} are arranged with respect to the $SU(3)$ and $U(1)$ 
subgroups of the group $SO(6)$ into a singlet, a triplet and a sextet. 
The corresponding  antidecuplet  manifests as an antisinglet, an antitriplet and an antisextet.  
${\bf \tau^{33}}= \frac{1}{2} ({\cal {\bf S}}^{9\,10} - {\cal {\bf S}}^{11\,12})$, 
${\bf \tau^{38}}=$ $\frac{1}{2 \sqrt{3}} ({\cal {\bf S}}^{9\,10} + {\cal {\bf S}}^{11\,12} - 
2 {\cal {\bf S}}^{13\,14})$,  ${\bf \tau^{4}}=$ - 
$\frac{1}{3} ({\cal {\bf S}}^{9\,10} + {\cal {\bf S}}^{11\,12} + {\cal {\bf S}}^{13\,14})$;
${\cal {\bf S}}^{a b}$ $= i (\theta^a \frac{\partial}{\partial \theta_b} - \theta^b 
\frac{\partial}{\partial \theta_a})$. 
}
 \end{table}
%
The "fermion" quantum number ${\bf \tau^{4}}$ has for either singlets or triplets in Grassmann 
space, Table~\ref{Table grasssextet.}, twice the value of the corresponding singlets and triplets
 in Clifford space, Table~\ref{Table so13+1.}: $(-1, +1)$ in 
Grassmann case to be compared with $(- \frac{1}{2}, +\frac{1}{2})$ in Clifford case and 
$(+ \frac{1}{3}, -\frac{1}{3})$ in  Grassmann 
case to be compared with $(+ \frac{1}{6}, -\frac{1}{6})$ in Clifford case. 

\begin{figure}
  \centering
   \includegraphics[width=0.45\textwidth]{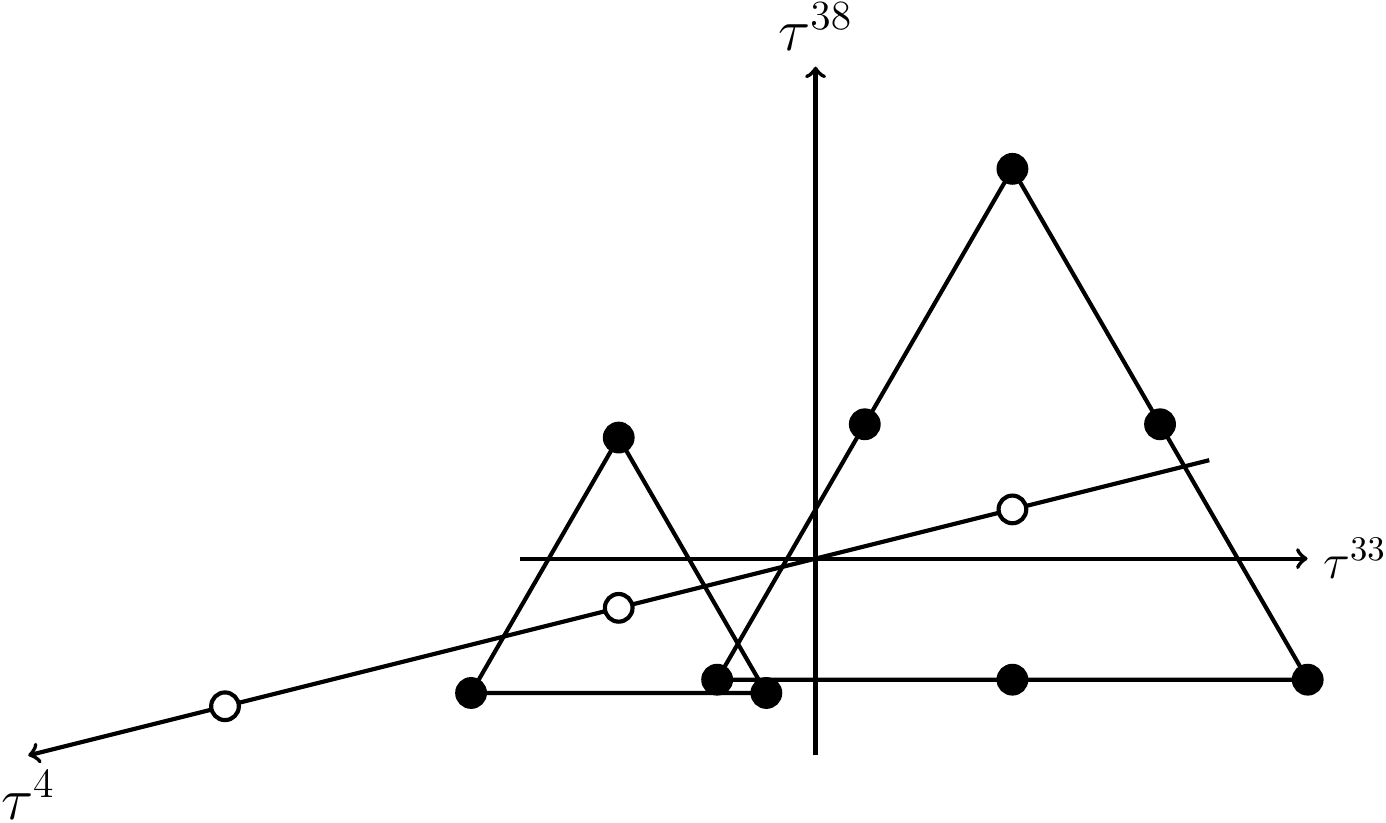} 
   \hfill
   \includegraphics[width=0.45\textwidth]{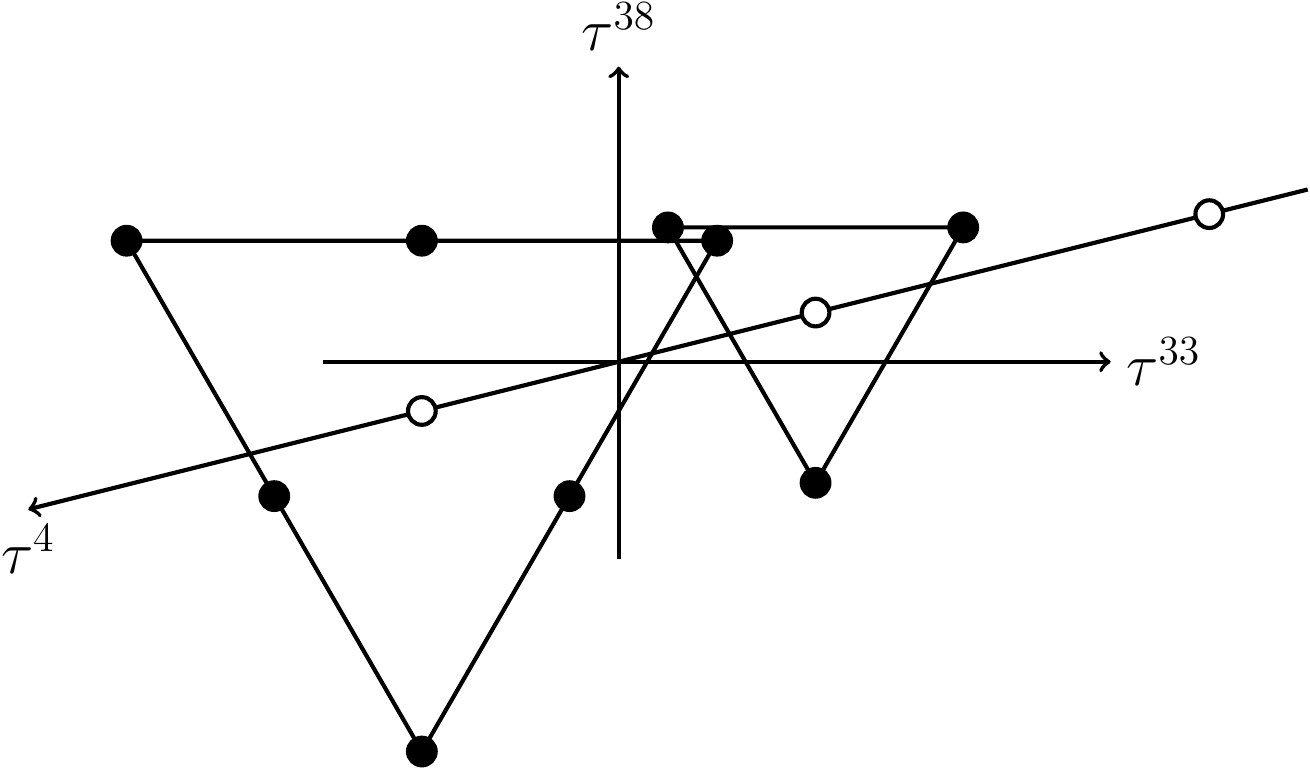}
  \caption{\label{FigSO6} Representations of the subgroups $SU(3)$ and $U(1)$ of the group 
$SO(6)$ in Grassmann space for two Grassmann odd representations of 
Table~\ref{Table grasssextet.} are presented. 
On the abscissa axis and on the ordinate axis the values of the two diagonal operators, 
${\bf \tau^{33}}$ and ${\bf \tau^{38}}$ of the coulour ($SU(3)$)
subgroup are presented, respectively, with full circles. On the third axis the values of the 
subgroup of the "fermion number" $U(1)$ is presented with the open circles, the same for all 
the representations of each  multiplet. There are one singlet, one triplet and one sextet on the 
left hand side and one antisinglet, one antitriplet and one antisextet on the right hand side.} 
\end{figure}
%

When  $SO(6)$ is embedded into $SO(13,1)$, the $SO(6)$ representations of either even or odd 
Grassmann character contribute to both  of the decoupled, 
$1716$ states of $SO(13,1)$ representations
contribute, provided that the $SO(8)$ content has the opposite Grassmann character than the $SO(6)$
content. The product of both representations must be Grassmann odd in order that the corresponding 
creation and annihilation operators fulfill the required anticommutation relations for fermions, 
Eq.~(\ref{ijthetaprod}).
\subsubsection{Properties of the subgroups $SO(3,1)$ and $SO(4)$ of the group $SO(8)$ in 
Grassmann and in Clifford space, when  $SO(8)$ is embedded into  $SO(13,1)$}
\label{so8}

{\bf a.}
Let us again repeat first properties of the $SO(3,1)$ and $SO(4)$ parts of the $SO(13,1)$
representation of  $64$ "family members" in Clifford space, presented in Table~\ref{Table so13+1.}. 
As seen in Table~\ref{Table so13+1.} there are four octets and four antioctets of $SO(8)$. 
All four octets, having an even Clifford character and forming $32$ states when embedded into 
$SO(13,1)$, are the same for either
quarks or for leptons, they distinguish only in the $SO(6)$ part (of a Clifford odd character) of the 
$SO(13,1)$ group, that is in the colour ($SU(3)$) part and  the "fermion quantum number" 
($U(1)$) part. 
Also the four antioctets, having an odd Clifford character, are all the same for the $32$ 
family members of antiquarks and antileptons, they again distinguish only in the Clifford 
even $SO(6)$ part of $SO(13,1)$, that is in the anticolour ($SU(3)$) part and the "fermion 
quantum number" ($U(1)$) part.

The $64^{th}$-plet of creation operators has an odd Clifford character either for quarks and 
leptons  or for antiquarks and antileptons --- correspondingly have an odd 
Clifford character also their annihilation operators --- and can be  second quantized~\cite{nh2018}.

Let us analyze first the octet ($2^{\frac{8}{2}-1}=8$), which  is the same for all $32$ members of
quarks and leptons. The octet has an even Clifford character. All the right handed $u_{R}$-quarks 
and  $\nu_{R}$-leptons have the $SO(4)$ part of $SO(8)$ equal to
$\stackrel{56}{[+]}\,\stackrel{78}{(+)}$, while their left handed partners have  the $SO(4)$ part of 
$SO(8)$ equal to $\stackrel{56}{[+]}\,\stackrel{78}{[-]}$.
All the right handed $d_{R}$-quarks  and  $e_{R}$-leptons have the $SO(4)$ part of $SO(8)$ 
equal to $\stackrel{56}{(-)}\,\stackrel{78}{[-]}$, while their left handed partners have the $SO(4)$ 
part of $SO(8)$ equal to $\stackrel{56}{(-)}\,\stackrel{78}{(+)]}$. The left handed quarks and 
leptons are doublets with respect to $\vec{\tau}^{1}$ and singlets with $\vec{\tau}^{2}$, while 
the right handed quarks and leptons are  singlets with respect to $\vec{\tau}^{1}$ and doublets
 with $\vec{\tau}^{2}$. 
The left and right handed quarks and lepton belong with respect to the $SO(3,1)$ group to either
left handed or the right handed spinor representations, respectively.

{\bf b.}
In Grassmann space the $SO(8)$ group of an odd Grassmann character has $\frac{1}{2}$ 
$\frac{8!}{4! 4!} = 35$ creation operators in each of the two groups and the same number of 
annihilation operators, obtained from the creation operators by Hermitian conjugation, 
Eq.~(\ref{grassher}). The corresponding states, created by the creation operators on the vacuum 
state $|\phi_{o}>$, can be therefore second quantized. But if embedded the group $SO(8)$ into
the group $SO(13,1)$ the subgroup $SO(6)$ must have an even Grassmann character in oder 
that the states in $SO(13,1)$ can be second quantized according to Eq.~(\ref{ijthetaprod}). 

According to what we learned in the case of the group $SO(6)$, each of the two independent 
representations of the group $SO(13,1)$ of an odd Grassmann character must   
include either the even $SO(7,1)$ part and the odd $SO(6)$ part or  the odd 
$SO(7,1)$ part and the even $SO(6)$ part. To the even $SO(7,1)$ representation 
either the odd $SO(3,1)$  and the odd $SO(4)$  parts  contribute or both must be of the 
Grassmann even character. In the case that the $SO(7,1)$ part has an odd Grassmann character 
(in this case the $SO(6)$ has an even Grassmann character) then one of the
two parts $SO(3,1)$  and $SO(4)$ must be odd and the other even.

\section{Concluding remarks}
\label{conclusions}

We learned in this contribution that although  either Grassmann or Clifford space offer the second
quantizable description of the internal degrees of freedom of fermions (Eq.~(\ref{ijthetaprod})),
the Clifford space offers more: It offers  not only the description of all the "family members", 
explaining all the degrees of freedom of the observed quarks and leptons and antiquark and antileptons, 
but also the explanation for the appearance of families. 

The interaction of fermions with the gravity fields --- the vielbeins and the spin connections
 --- in the $2(2n+1)$-dimensional space can be achieved, as suggested by the {\it spin-charge-family}
theory~(\cite{normaJMP2015,n2014matterantimatter} and references therein), by replacing the 
momentum $p_{a}$ in the Lagrange density function for a free particle by the covariant momentum,  
equally appropriate for both representations. In Grassmann space we have: $p_{0a}= f^{\alpha}{}_a$
 $p_{0\alpha}$, with $p_{0\alpha} = p_{\alpha}  - \frac{1}{2}\, {\cal {\bf S}}^{ab} 
\Omega_{ab \alpha}$, where $ f^{\alpha}{}_a$ is the vielbein in $d=2(2n+1)$-dimensional space and
$\Omega_{ab \alpha}$ is the spin connection field of the Lorentz generators ${\cal {\bf S}}^{ab}$.
In Clifford space we have equivalently: $p_{0a}= f^{\alpha}{}_a$  $p_{0\alpha}$,  $p_{0\alpha}= 
 p_{\alpha}  -  \frac{1}{2}  S^{ab} \omega_{ab \alpha} - \frac{1}{2}  \tilde{S}^{ab}  
 \tilde{\omega}_{ab \alpha}$. Since ${\cal {\bf S}}^{ab} = S^{ab}  + \tilde{S}^{ab}$ we find that
when no fermions are present either $\Omega_{ab \alpha}$ or $\omega_{ab \alpha}$ or 
$\tilde{\omega}_{ab \alpha}$ are uniquely expressible by vielbeins $f^{\alpha}{}_a$
 (\cite{normaJMP2015,n2014matterantimatter} and references therein). It might be that
 "our universe made a choice between the Clifford and the Grassmann algebra" when breaking
 the starting symmetry by making condensates of fermions, since that for breaking symmetries  
Clifford space offers better opportunity".
\appendix 
\section{Useful relations 
in Grassmann and Clifford space}
\label{grassmannandcliffordfermions}

The generator of the Lorentz transformation in Grassmann space is defined as follows~\cite{norma93}
\begin{eqnarray}
\label{Lorentztheta}
{\cal {\bf S}}^{ab} &=&  (\theta^a p^{\theta b} - \theta^b p^{\theta a})\,\nonumber\\
&=& S^{ab} +\tilde{S}^{ab} \,, \quad  \{S^{ab}, \tilde{S}^{cd}\}_{-} =0\,,
\end{eqnarray}
where $S^{ab}$ and $\tilde{S}^{ab}$ are the corresponding two generators of the Lorentz
 transformations in the Clifford space, forming orthogonal representations with respect to each other.

We make a choice of the Cartan subalgebra of the Lorentz algebra as follows 
\begin{eqnarray}
&& {\cal {\bf S}}^{03}, {\cal {\bf S}}^{12}, {\cal {\bf S}}^{56}, \cdots, 
{\cal {\bf S}}^{d-1\; d}\,, \nonumber\\
&& S^{03}, S^{12}, S^{56}, \cdots, S^{d-1\; d}\,,\nonumber\\
&& \tilde{S}^{03}, \tilde{S}^{12}, \tilde{S}^{56}, \cdots, \tilde{S}^{d-1\; d}\,,\nonumber\\
 &&{\rm if } \quad d = 2n\,.
\label{choicecartan}
\end{eqnarray}
We find the infinitesimal generators of the Lorentz transformations in 
 Clifford space
\begin{eqnarray}
\label{Lorentzgammatilde}
S^{ab} &=& \frac{i}{4} (\gamma^a \gamma^b - \gamma^b \gamma^a)\,, \quad
S^{ab \dagger} = \eta^{aa} \eta^{bb} S^{ab}\,,\nonumber\\
\tilde{S}^{ab} &=& \frac{i}{4} (\tilde{\gamma}^a \tilde{\gamma}^b -
\tilde{\gamma}^b\tilde{\gamma}^a) \,,  \quad \tilde{S}^{ab \dagger} =
\eta^{aa} \eta^{bb} \tilde{S}^{ab}\,,
\end{eqnarray}
where $\gamma^a$ and $\tilde{\gamma}^a$ are defined in Eq.~(\ref{cliffthetarel}). 
The commutation relations for either ${\cal {\bf S}}^{ab}$ or $S^{ab}$ or $\tilde{S}^{ab}$,
${\cal {\bf S}}^{ab} = S^{ab} + \tilde{S}^{ab}$, are 
%
%
%
\begin{eqnarray}
\label{Lorentzthetacom}
\{S^{ab}, \tilde{S}^{cd}\}_{-}&=& 0\,, \nonumber\\
\{S^{ab},S^{cd}\}_{-} &=& i (\eta^{ad} S^{bc} + \eta^{bc} S^{ad} -
 \eta^{ac} S^{bd} - \eta^{bd} S^{ac})\,,\nonumber\\
\{\tilde{S}^{ab},\tilde{S}^{cd}\}_{-} &=& i(\eta^{ad} \tilde{S}^{bc} + 
\eta^{bc} \tilde{S}^{ad} 
- \eta^{ac} \tilde{S}^{bd} - \eta^{bd} \tilde{S}^{ac})\,.
\end{eqnarray}
The infinitesimal generators of the two invariant subgroups of the group $SO(3,1)$ can be expressed 
as follows
\begin{eqnarray}
\label{so1+3}
\vec{N}_{\pm}(= \vec{N}_{(L,R)}): &=& \,\frac{1}{2} (S^{23}\pm i S^{01},S^{31}\pm i S^{02}, 
S^{12}\pm i S^{03} )\,.
\end{eqnarray}
The infinitesimal generators of the two invariant subgroups of the group $SO(4)$ are expressible with
$S^{ab}, (a,b) = (5,6,7,8)$ as follows 
 \begin{eqnarray}
 \label{so42}
 \vec{\tau}^{1}:&=&\frac{1}{2} (S^{58}-  S^{67}, \,S^{57} + S^{68}, \,S^{56}-  S^{78} )\,,
\nonumber\\
 \vec{\tau}^{2}:&=& \frac{1}{2} (S^{58}+  S^{67}, \,S^{57} - S^{68}, \,S^{56}+  S^{78} )\,,
 \end{eqnarray}
while the generators of the $SU(3)$ and  $U(1)$ subgroups of the group $SO(6)$ can be expressed by
$S^{ab}, (a,b) = (9,10,11,12,13,14)$
 \begin{eqnarray}
 \label{so64}
 \vec{\tau}^{3}: = &&\frac{1}{2} \,\{  S^{9\;12} - S^{10\;11} \,,
  S^{9\;11} + S^{10\;12} ,\, S^{9\;10} - S^{11\;12} ,\nonumber\\
 && S^{9\;14} -  S^{10\;13} ,\,  S^{9\;13} + S^{10\;14} \,,
  S^{11\;14} -  S^{12\;13}\,,\nonumber\\
 && S^{11\;13} +  S^{12\;14} ,\, 
 \frac{1}{\sqrt{3}} ( S^{9\;10} + S^{11\;12} - 
 2 S^{13\;14})\}\,,\nonumber\\
 \tau^{4}: = &&-\frac{1}{3}(S^{9\;10} + S^{11\;12} + S^{13\;14})\,.
 \end{eqnarray}
The hyper charge $Y$ can be defined as $Y=\tau^{23} + \tau^{4}$. 

The equivalent expressions for the "family" charges, expressed by $\tilde{S}^{ab}$ follow if in 
Eqs.~(\ref{so1+3} - \ref{so64}) $S^{ab}$ are replaced by $\tilde{S}^{ab}$.

The breaks of the symmetries, manifesting in Eqs.~(\ref{so1+3}, \ref{so42}, \ref{so64}), are in 
the {\it spin-charge-family} theory caused by the condensate and the nonzero vacuum expectation 
values (constant values) of the scalar fields carrying the space index $(7,8)$ 
(Refs.~\cite{normaJMP2015,IARD2016} and the references therein). The space breaks first to 
$SO(7,1)$
$\times SU(3) \times U(1)_{II}$ and then further to $SO(3,1)\times SU(2)_{I} \times U(1)_{I}$
$\times SU(3) \times U(1)_{II}$, what explains the connections between the weak and the hyper 
charges and the handedness of spinors.

Let ius present some useful relations~\cite{IARD2016}
\begin{eqnarray}
\stackrel{ab}{(k)}\stackrel{ab}{(k)}& =& 0\,, \quad \quad \stackrel{ab}{(k)}\stackrel{ab}{(-k)}
= \eta^{aa}  \stackrel{ab}{[k]}\,, \quad \stackrel{ab}{(-k)}\stackrel{ab}{(k)}=
\eta^{aa}   \stackrel{ab}{[-k]}\,,\quad
\stackrel{ab}{(-k)} \stackrel{ab}{(-k)} = 0\,, \nonumber\\
\stackrel{ab}{[k]}\stackrel{ab}{[k]}& =& \stackrel{ab}{[k]}\,, \quad \quad
\stackrel{ab}{[k]}\stackrel{ab}{[-k]}= 0\,, \;\;
\quad \quad  \quad \stackrel{ab}{[-k]}\stackrel{ab}{[k]}=0\,,
 \;\;\quad \quad \quad \quad \stackrel{ab}{[-k]}\stackrel{ab}{[-k]} = \stackrel{ab}{[-k]}\,,
 \nonumber\\
\stackrel{ab}{(k)}\stackrel{ab}{[k]}& =& 0\,,\quad \quad \quad \stackrel{ab}{[k]}\stackrel{ab}{(k)}
=  \stackrel{ab}{(k)}\,, \quad \quad \quad \stackrel{ab}{(-k)}\stackrel{ab}{[k]}=
 \stackrel{ab}{(-k)}\,,\quad \quad \quad 
\stackrel{ab}{(-k)}\stackrel{ab}{[-k]} = 0\,,
\nonumber\\
\stackrel{ab}{(k)}\stackrel{ab}{[-k]}& =&  \stackrel{ab}{(k)}\,,
\quad \quad \stackrel{ab}{[k]}\stackrel{ab}{(-k)} =0\,,  \quad \quad 
\quad \stackrel{ab}{[-k]}\stackrel{ab}{(k)}= 0\,, \quad \quad \quad \quad
\stackrel{ab}{[-k]}\stackrel{ab}{(-k)} = \stackrel{ab}{(-k)}\,.
\label{graphbinoms}
\end{eqnarray}

\section*{Acknowledgment}
The author N.S.M.B. thanks Department of Physics, FMF, University of Ljubljana, Society of 
Mathematicians, Physicists and Astronomers of Slovenia,  for supporting the research on the 
{\it spin-charge-family} theory by offering the room and computer facilities and Matja\v z 
Breskvar of Beyond Semiconductor for donations, in particular for the annual workshops entitled 
"What comes beyond the standard models". 


\begin{thebibliography}{99}
\bibitem{nh2018} N. S. Manko\v c Bor\v stnik,  H.B. Nielsen, "Why nature made a choice of 
            Clifford and not Grassmann coordinates",  Proceedings  to  the $20^{th}$ Workshop "What comes 
             beyond the standard models", Bled, 9-17 of July, 2017, Ed. N.S. Manko\v c Bor\v stnik, H.B. Nielsen,
              D. Lukman, DMFA  Zalo\v zni\v stvo, Ljubljana, December 2017, p. 89-120 
              [arXiv:1802.05554v2] [arXiv:1806.01629 whole proceedings].
\bibitem{norma93} N. Manko\v c Bor\v stnik, "Spinor and vector representations in four dimensional Grassmann
              space", {\it J. of Math. Phys.} {\bf 34} (1993), 3731-3745.
\bibitem{IARD2016} N.S. Manko\v c Bor\v stnik, "Spin-charge-family theory is offering next step
             in understanding elementary particles and fields and correspondingly universe", 
             Proceedings to the Conference on Cosmology, Gravitational Waves and Particles, 
             IARD conferences, Ljubljana, 6-9 June 2016, The $10^{th}$ Biennial Conference 
             on Classical and Quantum Relativistic Dynamics of Particles and Fields,  
                          J. Phys.: Conf. Ser. {\bf 845} 012017 
    [arXiv:1409.4981, arXiv:1607.01618v2].
\bibitem{n2014matterantimatter}  N.S. Manko\v c Bor\v stnik, "Matter-antimatter asymmetry in the 
{\it spin-charge-family} theory", {\it Phys. Rev.} {\bf D 91}  065004 (2015) [arXiv:1409.7791]. 
\bibitem{normaJMP2015} N.S. Manko\v c Bor\v stnik, "The explanation for the origin of the 
Higgs  scalar and for the Yukawa couplings by the {\it spin-charge-family} theory", 
{\it J.of Mod. Physics} {\bf 6} (2015) 2244-2274, http://dx.org./10.4236/jmp.2015.615230
              [arXiv:1409.4981].
\bibitem{JMP2013} N.S. Manko\v c Bor\v stnik N S, "The spin-charge-family theory is explaining the
 origin of families, of the Higgs and the Yukawa couplings", {\it J. of Modern Phys.} {\bf 4} (2013) 823
[arXiv:1312.1542].
\bibitem{nd2017} N.S. Manko\v c Bor\v stnik, D. Lukman, "Vector and scalar gauge fields with
              respect to $d=(3+1)$ in Kaluza-Klein theories and in the {\it spin-charge-family theory}",
              {\it Eur. Phys. J. C} {\bf 77} (2017) 231.
\bibitem{nh2017} N.S. Manko\v c Bor\v stnik, H.B.F. Nielsen, "The spin-charge-family theory 
             offers understanding of the triangle anomalies cancellation in the standard model",
             {\it Fortschritte der Physik, Progress of Physics} (2017) 1700046.
\bibitem{nh02}  N.S. Manko\v c Bor\v stnik, H.B.F. Nielsen, {\it J. of Math. Phys.} {\bf 43}, 
5782 (2002) [arXiv:hep-th/0111257].
\bibitem{nh03} N.S. Manko\v c Bor\v stnik, H.B.F. Nielsen,
{\it J. of Math. Phys.} {\bf 44} 4817 (2003) [arXiv:hep-th/0303224].
\bibitem{n2012scalars} N.S. Manko\v c Bor\v stnik, "The {\it spin-charge-family} theory explains  
              why  the scalar Higgs carries the weak charge $\pm \frac{1}{2}$ and the hyper charge 
              $ \mp \frac{1}{2}$",
              Proceedings to 
              the $17^{th}$ Workshop "What comes beyond the standard models", Bled, 20-28 of July, 2014, 
              Ed. N.S. Manko\v c Bor\v stnik, H.B. Nielsen, D. Lukman, DMFA  Zalo\v zni\v stvo, 
              Ljubljana December 2014, p.163-82 [arXiv:1502.06786v1, arXiv:1409.4981].  

\bibitem{portoroz03} A. Bor\v stnik Bra\v ci\v c, N.S. Manko\v c Bor\v stnik,``The approach 
Unifying Spins and Charges and Its Predictions'',
				Proceedings to the Euroconference on Symmetries Beyond the Standard Model'',
				Portoro\v z, July 12 - 17, 2003,
			        Ed. by N.S. Manko\v c Bor\v stnik, H.B. Nielsen, C. Froggatt, D. Lukman,
 DMFA Zalo\v zni\v stvo, Ljubljana December 2003, p. 31-57 [arXiv:hep-ph/0401043, 
arXiv:hep-ph/0401055].\\
 \bibitem{pikanorma} A. Bor\v stnik Bra\v ci\v c and N.S. Manko\v c Bor\v stnik,
 ``Origin of families of fermions and their mass matrices'',
 {\it Phys. Rev.} { \bf D 74}, 073013 (2006) [arXiv:hep-ph/0301029, arXiv:hep-ph/9905357, p. 
52-57, arXiv:hep-ph/0512062,  p.17-31, arXiv:hep-ph/0401043, p. 31-57].
\bibitem{DHN} D. Lukman, N.S. Manko\v c Bor\v stnik and H.B. Nielsen,
"An effective two dimensionality cases bring a new hope to the Kaluza-Klein-like theories", 
{\em New J. Phys.} 13:103027, 2011.
%
\bibitem{DN012} 
D. Lukman and N.S. Manko\v c Bor\v stnik,  
"Spinor states on a curved infinite disc  with non-zero spin-connection fields",
{\em J. Phys. A:  Math. Theor.} 45:465401, 2012 
[arXiv:1205.1714, arXiv:1312.541, arXiv:hep-ph/0412208 p.64-84].
\bibitem{familiesDNproc}  D. Lukman, N.S. Manko\v c Bor\v stnik and H.B. Nielsen,
  "Families of spinors in $d=(1+5)$ with a zweibein and two kinds of spin connection fields on 
 an almost $S^2$",  Proceedings 
               to the $15^{th}$ Workshop "What comes beyond the standard models", Bled, 
                9-19 of July, 2012, Ed. N.S. Manko\v c Bor\v stnik, 
                H.B. Nielsen, D. Lukman, DMFA  Zalo\v zni\v stvo, Ljubljana December 2012, 157-166
                [arXiv:1302.4305].
\bibitem{gn2009} G. Bregar and N.S. Manko\v c Bor\v stnik, "Does dark matter consist of baryons
 of new stable family quarks?", {\it Phys. Rev.} {\bf D 80},
 083534 (2009) 1-16.
\bibitem{pikan2006}  A. Bor\v stnik Bra\v ci\v c, N. S. Manko\v c Bor\v stnik, ''On the origin of families 
                     of fermions and their mass matrices'', [arXiv:hep-ph/0512062],  
Phys Rev. {\bf D 74} 
                     073013-28  (2006).
\bibitem{nhds}  N.S. Manko\v c Bor\v stnik, H.B. Nielsen, "Discrete 
                   symmetries in the Kaluza-Klein-like theories", 
                   {\it Jour. of High Energy Phys.} {\bf 04} (2014) 165
                   [arXiv:1212.2362v3].
%
\end{thebibliography}
\end{document}